\documentclass{article}

\usepackage[utf8]{inputenc}				
\usepackage[big,online]{dgruyter}	
\usepackage{lmodern} 
\usepackage{microtype}
\usepackage[numbers,square,sort&compress]{natbib}

\usepackage[ruled, linesnumbered]{Sty/algorithm2e}
\SetAlFnt{\small}
\SetAlCapFnt{\small}
\SetAlCapNameFnt{\small}
\usepackage{algorithmic}
\algsetup{linenosize=\tiny}

\usepackage{amsmath}
\usepackage{amssymb}
\usepackage{amsthm}
\usepackage{bm}
\usepackage{makecell}
\usepackage{graphicx}
\usepackage{tabularx}
\usepackage{Sty/bbm}
\usepackage{Sty/multirow}
\usepackage{Sty/mathtools}
\usepackage{Sty/graphicx}
\usepackage{url}
\usepackage{setspace}
\usepackage{pdflscape}

\begin{document}

\articletype{Preprint}

\title{Multi-agent statistical discriminative sub-trajectory mining and an application to NBA basketball}
\runningtitle{Multi-agent statistical discriminative trajectory mining}

\author*[1]{Rory Bunker}
\author[2]{Vo Nguyen Le Duy}
\author[3]{Yasuo Tabei}
\author[4]{Ichiro Takeuchi}
\author[5]{Keisuke Fujii}
\runningauthor{Bunker et al.}
\affil[1]{Graduate School of Informatics, Nagoya University, Email: rory.bunker@g.sp.m.is.nagoya-u.ac.jp}
\affil[2]{Data-Driven Biomedical Science Team,
RIKEN Center for Advanced Intelligence Project. Email: duy.vo@riken.jp}
\affil[3]{RIKEN Center for Advanced Intelligence Project. Email: yasuo.tabei@riken.jp}\affil[4]{Graduate School of Engineering Mechanical Systems Engineering, Nagoya University. Email: ichiro.takeuchi@mae.nagoya-u.ac.jp}
\affil[5]{Graduate School of Informatics, Nagoya University; RIKEN Center for Advanced Intelligence Project; PRESTO, Japan Science and Technology Agency. Email: fujii@i.nagoya-u.ac.jp}

\abstract{Improvements in tracking technology through optical and computer vision systems have enabled a greater understanding of the movement-based behaviour of multiple agents, including in team sports. In this study, a Multi-Agent Statistically Discriminative Sub-Trajectory Mining (MA-Stat-DSM) method is proposed that takes a set of binary-labelled agent trajectory matrices as input and incorporates Hausdorff distance to identify sub-matrices that statistically significantly discriminate between the two groups of labelled trajectory matrices. Utilizing 2015/16 SportVU NBA tracking data, agent trajectory matrices representing attacks consisting of the trajectories of five agents (the ball, shooter, last passer, shooter defender, and last passer defender), were truncated to correspond to the time interval following the receipt of the ball by the last passer, and labelled as effective or ineffective based on a definition of attack effectiveness that we devise in the current study. After identifying appropriate parameters for MA-Stat-DSM by iteratively applying it to all matches involving the two top- and two bottom-placed teams from the 2015/16 NBA season, the method was then applied to selected matches and could identify and visualize the portions of plays, e.g., involving passing, on-, and/or off-the-ball movements, which were most relevant in rendering attacks effective or ineffective.
}
  \keywords{team sports, trajectory analysis, tracking data, Hausdorff distance, geographic information systems, spatial information}
  \startpage{1}
  \aop

\maketitle
\vspace{-0pt}
\section{Introduction}
\vspace{-0 pt}

The development of tracking technology has increased the availability of trajectory data in various domains, and trajectory mining techniques have been developed and applied in various fields, e.g., in biology to understand animal behavior \citep{li2011movemine}, and to understand pedestrian movements \citep{cao2016algorithm}.
In team sports, tracking data, obtained from video, wearable devices, or optical systems, has traditionally been used primarily by strength and conditioning staff to analyze athlete movements and demands, e.g., to determine the optimal times to replace players during a match to maximize performance and minimize injuries \citep{carling2008role, karcher2014court}. 
However, spatiotemporal tracking data also has potential value for sport performance analysts, who can complement their usual analysis of performance indicators \citep{hughes2002use}, derived from event log data captured in video analysis systems such as SportsCode or Dartfish, with analysis derived from tracking data, for a more holistic understanding of player and team performance.
Both the review papers of \cite{rein2016big} and \cite{goes2021unlocking} highlighted the potential benefits of greater collaboration between sports scientists and computer scientists to explore greater use of spatiotemporal tracking data for performance analysis (in soccer but also team sports in general).
In basketball, matches can be decomposed into quarters that are, in turn, decomposed into individual plays.
There is value to coaches and performance analysts in identifying the most important parts of plays, e.g., the portions of plays that discriminate between effective and ineffective attacks.

Statistically Discriminative Sub-Trajectory Mining (Stat-DSM) \cite{le2020stat} is a sub-trajectory mining method (a type of trajectory mining \cite{zheng2015trajectory,mazimpaka2016trajectory} algorithm) that identifies sub-trajectories that statistically significantly discriminate between labeled groups of trajectories of a single agent (hereafter, statistically significantly discriminative is shortened to ``SSD'' or simply ``discriminative'').
As well as proposing the Stat-DSM method, \cite{le2020stat} demonstrated its applicability on datasets consisting of hurricane and vehicle trajectories.
The Stat-DSM method cannot be directly applied in team sports in general, because there are multiple trajectories corresponding to the movements of multiple agents (players and the ball).
Therefore, in the current study, we propose an extension of Stat-DSM, Multi-Agent (MA) Stat-DSM that aims to identify SSD \textit{sub-matrices}, which consist of the sub-trajectories of multiple agents.
To identify SSD sub-trajectories or SSD sub-matrices in the case of Stat-DSM or MA-Stat-DSM, respectively, a distance metric \cite{su2020survey} is required.
While Euclidean distance is used in Stat-DSM to compute the distance between two (sub)-trajectories of a single agent, in the multi-agent setting, the distance between matrices of agent trajectories, which are of differing lengths across attacks, needs to be determined.
Inspired by its use in the point set distance metric approach in multiple-instance learning \citep{Herrera2018foundations}, we incorporate Hausdorff distance in MA-Stat-DSM to determine the distance between (sub)-matrices of differing lengths.

%
%

To demonstrate the applicability of the proposed MA-Stat-DSM method, it is applied to the Stats Perform (Chicago, IL, USA) SportVU NBA optical tracking system data from the 2015/16 NBA basketball season, which was preprocessed to contain player and ball trajectories in attacks, which were sub-sampled at a frequency of 5Hz.
%
%
The trajectories of five agents --- two attacking players, two defending players, and the ball --- were considered.
In particular, each trajectory matrix represented an attack consisting of trajectories of the shooter, the last passer, the shooter defender (closest to the shooter at the time of the shot), and the last passer defender.
Each trajectory matrix was labeled based on whether it is an effective or ineffective play, with the effective and ineffective labels computed based on three factors: the position of the shooter on the court (court area/zone), the distance of the shooter from the nearest defender (whether the shot was wide open), and, in the case where a shot is attempted outside the circle, the shooter's historical shot success percentage.

The three main contributions of this study are as follows:
\begin{enumerate}
    \item A multi-agent statistical discriminative trajectory mining method, MA-Stat-DSM, is proposed that extends Stat-DSM to take the trajectories of multiple agents, in the form of a trajectory matrix, as input and identify the most relevant portion of each attack by obtaining SSD sub-matrices. 
    Unlike machine learning approaches, MA-Stat-DSM does not require complex feature engineering from point coordinates (e.g., by computing, velocities, accelerations, angles, etc.), and its underlying mechanisms are more intuitively understandable compared to black-box deep learning approaches. 
    \item A novel approach to defining effective and ineffective attacks in basketball is proposed based on the concept of wide-open shots. 
    Each attacking play (trajectory matrix) is labeled as effective or ineffective based on this definition.  
    \item The proposed method is demonstrated on SportVU NBA trajectory data.
    In particular, MA-Stat-DSM is applied to the attacks of a specific team in a particular match to identify the portions (sub-matrices) of attacks (agent trajectory matrices) that discriminate between effective and ineffective plays, which could reveal useful post-match insights for coaches and performance analysts.
\end{enumerate}

The remainder of the paper is organized as follows. 
An overview of related studies is provided in Section \ref{sec:related-work}.
Then, in Section \ref{sec:data}, we describe the trajectory dataset used in this study, including the proposed computation of effective and ineffective attack labels.
Section \ref{sec:method} then describes the proposed MA-Stat-DSM itself.
Section \ref{sec:results} provides visualizations of SSD sub-matrices in matches involving, for generality, top- and bottom-performing teams from the 2015/16 season.
Finally, section \ref{sec:discussion} discusses the obtained results, potential limitations, and avenues for further research.

\section{Related Work}
\label{sec:related-work}

Many basketball studies related to tracking data have utilized optical tracking data from the SportVU arena camera system of STATS Perform (prior to 2017 STATS provided tracking data to the NBA) derived from video footage obtained by multiple cameras in the basketball arenas \cite{terner2021modeling}.
As mentioned in the introduction, we also use the SportVU data, which is described further in subsection \ref{ssec:trajectory-data}.

Statistical methods, e.g., cluster analysis and analysis of variance (ANOVA), have been applied to tracking and non-tracking data to construct performance indicators and profiles related to scoring, passing, defensive and all-round game roles and behavior.
Network-based models \cite{skinner2015method} have also been applied to tracking data to enable the enhanced evaluation of individual player skills/performance and prediction of team performance in basketball that can surpass traditional statistics-based approaches.
Deep learning-based computer vision techniques have also been proposed to analyze tracking data in basketball, e.g., to classify player and ball movements from video and to analyze passing relationships \cite{yoon2019analyzing}.
Lucey et al. \cite{lucey2014get} used tracking data in the (3-second) lead-up to three-point attempts, to analyze movement patterns that create ``open shots'' (where the nearest defender to the shooter is at a distance of at least 6 feet away) and how these can impact performance.
As mentioned, the concept of open shots comprises part of the effective/ineffective attack label definition that is proposed in the current study.
Tracking data in the lead-up to three-point shots has been converted into sequences, and recurrent neural networks, a deep learning model, have been used to predict three-point shot success/failure \cite{shah2016applying}.
A Long-Term Short Term Memory Network
(LSTM) \cite{hochreiter1997long} with neural embedding and deep feature representation was proposed by \cite{sicilia2019deephoops}, who formulated a multi-class sequence classification problem that uses spatiotemporal tracking data as input.
The approach estimates the probabilities of actions taken by players at the end of possessions, which can determine expected points at each point in time during an attacking play and can, in turn, be used to evaluate so-called micro-actions in terms of their contribution to the success of a possession.

Strategy identification and classification in basketball is another area of study that uses tracking data.
For instance, Wang \& Zemel \cite{wang2016classifying} used neural and recurrent neural networks, which are able to handle sequential data that are of varying lengths, to SportVU trajectories that were converted into image representations for classification of attacking plays and sequence prediction, and to investigate whether the model could classify offensive plays in a subsequent season. 
One of the primary contemporary offensive strategies used in the NBA is the pick-and-roll/ball screen.
McQueen, Wiens, \& Guttag \cite{mcqueen2014automatically} applied a machine learning classification model on top of a rule-based algorithm to identify on-ball screens from SportVU tracking data from 21 quarters across 14 matches in the 2012/13 NBA season.
Building on this work, McIntyre et al. \cite{mcintyre2016recognizing} applied a supervised machine learning classifier to SportVU data from the 2011/12 to 2014/15 seasons to automatically recognize defensive strategies employed against ball screens.
Machine learning techniques, e.g., k-nearest neighbors, decision trees, and support vector machines, have been applied to various tracking data-derived features such as player velocities, inter-player distances, player movement vector similarity, and defensive zones, to classify defensive (switch and trap) strategies used against pick-and-rolls \cite{tian2019use}.
Using player tracking system data from 1,230 regular season matches in the 2013-2014 season, \cite{sampaio2015exploring} used discriminant analysis to distinguish between the performance of all-star and non all-star players in NBA basketball, and identified role-based performance profiles of players using k-means clustering.
Active learning with neural networks is another approach that has been proposed to circumvent time-consuming annotation by domain experts to identify the pick-and-roll offensive strategy using tracking data \cite{ai2021novel}.
Semi-supervised learning has also recently been proposed for the classification of ball screen plays from SportVU tracking data \cite{ziyi2022cooperative}.

Methods that track player and ball movements can be useful for performance evaluation and strategy identification/classification, however, it is not possible to holistically analyze team performance without considering the movements and interactions of all players as a group \cite{gudmundsson2017spatio}.
Multi-agent methods that consider the movements of agents including the players and ball are, therefore, important in this context, and deep learning approaches such as bidirectional LSTM and mixture density networks have been used for trajectory prediction \cite{zhao2018applying} and assisting in decision-making regarding the optimal locations and times to make a shot.
Tensors  \cite{papalexakis2018thoops} and transformers \cite{alcorn2021baller2vec} are other deep learning approaches that have been used to model multi-agent spatiotemporal data in basketball, and graph-based representations \cite{raabe2023graph} have also been proposed in other sports (soccer).

Of relevance to the current study are methods that identify relevant parts of plays, e.g., those that discriminate between good and bad outcomes or different types of periods of play.
For instance, Facchinetti et al. \cite{facchinetti2023filtering} developed an algorithm that discriminates between active and inactive periods of play using trajectory data derived from sensor tracking data.
Chen et al. \cite{chen2015spatio} converted video clip data into player trajectory/action representations to analyze offensive strategies of differing duration in basketball, using dynamic time warping to compute the similarity of video clips, and largely unsupervised approach with clustering used to divide training data and Gaussian mixture regression employed to robustly model discriminative between-label variations.
%
A multi-agent neural network-based approach based on an attention mechanism using features related to multi-agent movements e.g., the distances between agents and objects, was recently proposed to identify trajectory segments that are correlated with effective/ineffective and scoring/non-scoring attacks \cite{ziyi2023multi}.
The method proposed in the current study has two main advantages over \cite{ziyi2023multi}: first, it does not require the extraction of movement-related features from the original trajectory data, and second, it is more intuitive compared to the black-box nature of deep learning methods.
Discriminative methods have also been used in other sports to identify discriminative patterns from event sequence data.
For example, discriminative sequential pattern mining has been applied to event sequences derived from event log data in Rugby to identify subsequences (patterns) that discriminate between scoring and non-scoring plays \cite{bunker2021supervised}.
Interestingness measures from mined frequent sequential patterns have also been obtained during training in cycling \cite{hrovat2015interestingness}.

\section{Data}
\label{sec:data}
In this section, the preprocessed SportVU NBA trajectory dataset, to which the proposed method will be applied, is described in subsection \ref{ssec:trajectory-data}.
Then, in subsection \ref{ssec:label-definition}, the approach for the computation of the effective and ineffective attack labels is described.

\subsection{Trajectory data}
\label{ssec:trajectory-data}
In this study, we used attack sequences from 600 regular season games from the 2015/2016 NBA season, which was originally sourced from GitHub from the 2015-2016 NBA Raw SportVU Game Logs (\url{https://github.com/neilmj/BasketballData/tree/master/2016.NBA.Raw.SportVU.Game.Logs}).
The dataset originally contained the trajectories of 11 agents, five players on each opposing team and the ball.
We considered five of these agents, the ball, and two players from each opposing team: the shooter, the shooter defender, the last passer, and the defender of the last passer.

Since scoring prediction is generally difficult, and non-linear data-driven approaches may sometimes not be interpretable (e.g., \cite{Fujii17,Fujii18}), we defined the label to be based on whether or not a particular play was an ``effective attack'' rather than whether or not points were scored in that play.
The definition/computation of effective and ineffective attacks will be provided in detail in subsection \ref{ssec:label-definition}.

There were a total of 45,307 attacks, sub-sampled at 5 Hz.
Although the raw data is spatiotemporal tracking data, in the trajectories and sub-trajectories (and matrices and sub-matrices comprised of them), since the time between consecutive point coordinates is fixed, the agent coordinates are temporally aligned and, therefore, given the trajectories are temporally ordered, only need to consider the spatial aspect of the data.
In our dataset, there were 18,021 shot successes, 20,155 shot failures, 22,159 effective attacks, and 23,148 ineffective attacks.
This dataset was already split into attacks, but as a preprocessing step, we removed the duplicate attacks and trimmed the start and end times.
The probabilities of scoring, given the attack was effective and ineffective, were 0.466 and 0.333, respectively.


\subsection{Effective and ineffective attack labeling}
\label{ssec:label-definition}
In this subsection, we describe our approach to computing effective and ineffective attacks, which are used as the trajectory matrix (attack) labels.
Due to the differing shooting abilities of individual players and other stochastic factors, evaluating team movements based on scores alone may not provide a holistic view of a ``good'' attacking play.
%
%
Indeed, it could be argued that the tactics and strategy of a coach and team are most influential up until the point at which there is a good scoring opportunity, i.e., the creation of a chance to attempt a shot.
It is then the skills and form of the individual player that determines whether this scoring opportunity is actually converted into points.
We consider a good scoring opportunity in basketball to be a shot that is attempted in a context in which there is a high expected probability of scoring, based on historical attempted and successful shots.
Therefore, we compute an interpretable and simple indicator from available statistics, based on frequencies, to evaluate whether a player makes an effective shot attempt, rather than based on a label with only successful shots or learning-based score prediction.

From the available statistics, we focused on two basic factors for effective attacks at an individual player level: shot zone on the court, and the distance between a shooter and the nearest defender.
These two factors are considered to be important for basketball shot prediction \citep{Fujii16,Fujii17,Fujii18}.
Metulini et al. \cite{metulini2018modelling} assessed the impact of spacing among players on team performance using tracking data.
In the NBA advanced stats (\url{https://www.nba.com/stats/players/shots-closest-defender/}), we have access to the probabilities of successful shots attempted in each zone, as well as distances, for each player.
The shot zones are partitioned into four areas: restricted, in-the-paint, mid-range, and 3-point areas.
%
The restricted area is defined as the area within a radius of 2.44 m (the distance between the side of the rectangle and the hoop) from the hoop.
The in-the-paint area is defined as the area within a radius of 5.46 m (the distance between the hoop and the farthest vertex of the rectangle) from the hoop.
The three-point area is defined as the area that is outside of the 3-point line. 
The mid-range is the remaining area.
The distance of the shooter from the nearest defender is categorized into four ranges: $0-2$ feet, $2-4$ feet, $4-6$ feet, and $6+$ feet.

We define an effective attack as one that meets the following criteria:
\begin{quote}
\begin{itemize}
\item The position of the shooter in the restricted area is effective at any distance because of their proximity to the hoop (despite a defender often being located near the shooter).
\item The position of the shooter in-the-paint and mid-range areas is effective at a distance of six feet or more from the nearest defender (this range is regarded as ``open'' in the NBA advanced stats).
\item The position of the shooter in the 3-point area is effective when a player with a shot success probability of at least 0.35 attempts a shot at a distance of 6 feet or more from the nearest defender (because some players do not shoot tactically).
\end{itemize}
\end{quote}

The probability of 0.35 is determined by the simple idea of a 3-point shot being ``not bad.'' 
If we assume that 50\% of 2-point shots are successful (this is determined subjectively, but is not unrealistic), we can select a 3-point shot if more than 33\% of 3-point shots are successful. 
Therefore, we determined the threshold as 35\%. 
Of course, this is a rough estimation and ideally, this should be customized for each team's strategy, but this is beyond the scope of the current study.

Based on the statistics in the 2014/2015 season and the tracking data, we computed the probabilities of successful shots for each zone and the distances for each player.
We computed the probabilities of players who had attempted less than 10 shots based on those of players of the same position (i.e., guard, forward, center, guard/forward, and forward/center, based on the registration in the NBA 2014/2015 season).
It should be noted that certain characteristics of a good shot can differ depending on the court location and context, e.g., for 2- and 3-pointers.
Note that, unfortunately, we could access those for only two areas (the 2- and 3-point areas) with four distance categories.
Thus, we computed the shot success probabilities in the restricted, in-the-paint, and mid-range areas using those in the 2-point area.
To adjust the 3-point shot probabilities for shots attempted a long distance from the 3-point line, we linearly reduced the probabilities by 0.2 at 12.73 m (the distance between the hoop and the half-court line).
In a naïve approach, the results obtained using trajectories consisting of the entire attack segments would not be interpretable and would not provide useful results because the roles of the players are not aligned with the order of the players in the trajectory data.
To extract meaningful information, we focus on the trajectories of five agents (the ball and four players) in the interval from when the last passer receives the ball until the shooter makes a shot (interval t2 to t0 shown in Figure 1). 
%
%
When a shot is not attempted in a particular play, the end time of the trajectory is determined as the time at which a turnover occurs (i.e., when the defensive team comes to be in possession of the ball).

\begin{figure}[h]
\centering
\includegraphics[width=1\textwidth]{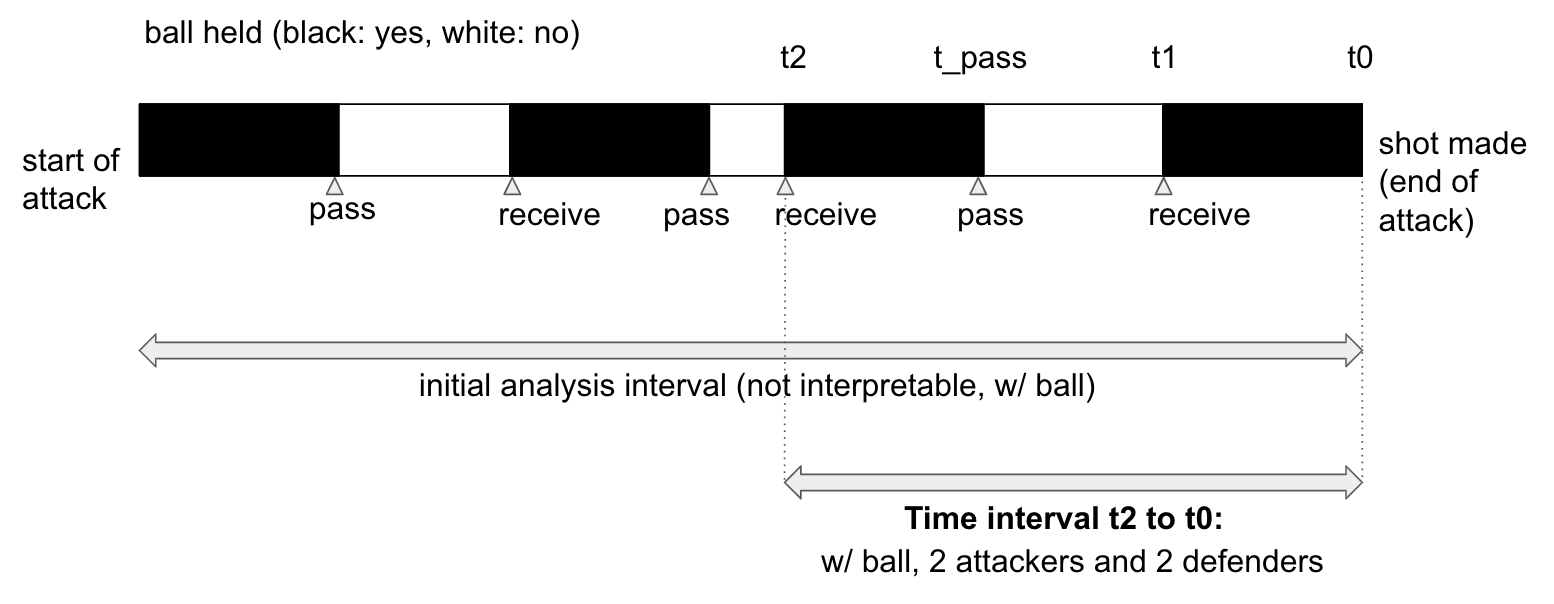}

\caption{The attacks were cropped to consider the time interval from t2 to t0, i.e., the trajectories of the ball, 2 attackers (shooter and last passer), and 2 defenders (shooter defender, last passer defender) during the time from which the last passer receives the ball until a shot is made (t1 is the time at which the shooter receives the ball).}
\label{fig:time_intervals}
\end{figure}

\section{Method}
\label{sec:method}
In this section, we first describe in subsection \ref{ssec:role-assignment} the benefits of applying MA-Stat-DSM to trajectory matrices (attacks) over Stat-DSM to individual agent trajectories, and how we resolve the resulting role assignment problem.
Then, we outline our problem setup and provide relevant definitions in subsections \ref{ssec:problem-setup} and \ref{ssec:definitions}, respectively.
Finally, we provide the pseudocode for the MA-Stat-DSM algorithm in \ref{ssec:algorithm}.

\subsection{Role Assignment}
\label{ssec:role-assignment}
One naïve approach would be to apply Stat-DSM to the trajectories of each agent.
However, this approach would result in losing potentially important spatiotemporal information about multi-agent interactions and would also require more computational time.
Furthermore, there may be interactions between players that would not be captured if the Stat-DSM method was applied to each individual agent trajectory and the results were combined.
Therefore, we apply MA-Stat-DSM to the set of multi-agent trajectories, which requires a less-than-proportional increase in computational time with respect to the number of agents considered. 

Consequently, a role-assignment problem occurs in multi-agent data processing.
Since the players are usually ordered randomly in the raw data, meaningful roles such as position (e.g., guard, forward, and center) are ignored.
Generally, this problem can be solved by a linear assignment problem \citep{Papadimitriou82}, e.g., using a Gaussian Hidden Markov Model \citep{le2017coordinated,fujii2020policy}.
However, since this approach is data-driven, it lacks the interpretability of each assigned role.
In this paper, we assign meaningful roles (e.g., shooter, shooter defender, etc.) in a rule-based manner to retain the interpretability of the obtained results.
In particular, in addition to the computational cost, the interpretability of the role is important for us, which is considered a role assignment problem, as mentioned.
We selected the five agents (ball, shooter, shooter defender, last passer, and last passer defender) because considering all trajectories may be too diverse for the model to extract useful information.
In general, this is a multi-agent role assignment problem for an unsorted, diverse dataset, which can be avoided by using only predetermined roles about these four players and the ball.
As mentioned, a data-driven approach was also considered, but we prioritized interpretability).
It is more difficult to determine the roles in a fixed manner as the number of players increases (e.g., it is difficult to generally determine the roles other than those of the shooter and passer), and fewer players may be less informative in this analysis.
Therefore, we consider that all trajectories with all player role assignments will lose some information in turn; thus, we selected these five agents. 

\subsection{Problem Setup}
\label{ssec:problem-setup}
The key differences between the proposed MA-Stat-DSM method and the original Stat-DSM method are that the labelled item in Stat-DSM is a trajectory whereas it is a trajectory matrix in MA-Stat-DSM, and instead of aiming to identify SSD sub-trajectories as in Stat-DSM, in MA-Stat-DSM, we aim to identify SSD sub-matrices (which are comprised of agent sub-trajectories).
Euclidean distance cannot be used to compute the distance between matrices that consist of multi-agent trajectories, so we incorporate an efficient implementation of Hausdorff distance \citep{huttenlocher1993comparing}, which was proposed by \citep{taha2015efficient}, and is available on GitHub (\url{https://github.com/mavillan/py-hausdorff}).

There is assumed to be a set of $K$ agents, each of which has a trajectory in all $N$ matrices.
Each trajectory matrix has $K$ rows and a length corresponding to the number of coordinates in each agent trajectory.
The lengths of each of the agent trajectories are the same within a trajectory matrix.
The length of the agent trajectories represents the number of columns in the trajectory matrix.
Therefore, the lengths of each of the agent trajectories being the same within a trajectory matrix implies that there are no point coordinates that have values.
On the other hand, different trajectory matrices will generally be of different lengths, i.e., have different numbers of columns.

In the context of the current study related to basketball, the agents represent the ball or a player, and the trajectory matrices represent plays.
In particular, there are $|K| = 5$ agents: $K$ = \{ball, shooter, shooter defender, last passer, last passer defender\}.
Each of the $K$ agents has a trajectory that is of the same length in each trajectory matrix (play).
A single agent in the set of all agents is denoted by $k \in K$.
The $i$th play/trajectory matrix, of which there are $N$ in total, consists of the trajectories of each agent in that play and can be represented as a $|K| \times m_i$ matrix, where $m_i$ is the length of play $i$.
In general, the lengths of each play differ, i.e., $m_i \neq m_j$ for $i \neq j$.

In this study, we apply the MA-Stat-DSM method to the attacks of a particular team in a specific match.
The method was initially applied to the attacks of a particular team across the whole 2015/16 season, but the computational complexity was found to be prohibitive, therefore, we considered instead the attacks of teams in a particular match and then iterated over the matches in the season.
Therefore, we aimed to use MA-Stat-DSM to identify SSD sub-matrices, i.e., portions of attacks, which discriminated between team $T$'s effective and ineffective attacks in match $M$.

\subsection{Definitions}
\label{ssec:definitions}
In this subsection, we provide some definitions and notation for Hausdorff distance, trajectory matrices, sub-matrices, distance, $\varepsilon$-neighborhood, and support, partly based on the definitions provided by \cite{zhao2018rest} and \cite{le2020stat}.
%

\textbf{Hausdorff distance.} 
%
The Hausdorff distance between two sets of instances (trajectory matrices in our case) is the aggregation of the base distances between the instances in each matrix.
Euclidean distance, Manhattan, or Chebyschev distance are commonly used as base distances (we used Euclidean distance in this study).
Two matrices, $X$ and $Y$, are within a Hausdorff distance of $dist_H$ if and only if every point in $X$ is within distance $dist_H$ of at least
one point in $Y$, and every point in $Y$ is within distance $dist_H$
of at least one point in $X$.
In particular, the Hausdorff distance, $dist_H(X,Y)$, between two point sets (matrices) is, in general: 
\begin{equation}
dist_H(X,Y) = \max\{h(X,Y),h(Y,X)\},
\label{eq:hausdorff-dist}
\end{equation}
where $h(X,Y) = $ \(\max\limits_{x \in X}\) \( \min\limits_{y \in Y}\) $dist(x,y)$, and 
 $h(Y,X) = $ \(\max\limits_{y \in Y}\) \( \min\limits_{x \in X}\) $dist(y,x)$.

\textbf{Trajectory matrix.}
%
%
The  trajectory of agent $k$ in matrix $i$ is a finite sequence of $m_i$ points: $T_{i,k} = \{(x_{1,k}, y_{1,k}), (x_{2,k}, y_{2,k}), ..., (x_{m_i,k}, y_{m_i,k})\}$.
Trajectory matrix $i$ contains the trajectories of all $K$ agents in a specific play, has $m_i$ columns, and is denoted $\bm T_{i,K}$.
%
%
There are $N$ trajectory matrices in a specific match $M$, each of which takes a label from $g_i = \{+1,-1\}$, and $G_+ = \{\bm T_{i,K} \mid g_i = +1\}$ and $G_- = \{\bm T_{i,K} \mid g_i = -1\}$ denote the groups of trajectory matrices with labels +1 and -1, respectively.
In the current study, as mentioned, a trajectory matrix represents an attack, which is labeled as either effective or ineffective.

\textbf{Trajectory sub-matrix.}
A sub-matrix is denoted $\bm T^{(s,e)}_{i,K}$, and is a sequence of consecutive columns within the trajectory matrix $\bm T_{i,k}$, starting from column index $s$ and ending at $e$, with a fixed number of $|K|$ rows. 
The length of sub-matrix $i$ is |$\bm T^{(s,e)}_{i,K}| \ge L$, where $L$, the minimum length (number of columns) of the sub-matrix, is the ``minimum length'' user-selected parameter of MA-Stat-DSM.
The notation $\bm T^{(s,e)}_{i,K} \sqsubseteq \bm T_{i,K}$ indicates that 
$\bm T^{(s,e)}_{i,K}$ is a sub-matrix of 
$\bm T_{i,K}$.
%

\textbf{Distance metric between sub-matrices.}
Using the general definition of Hausdorff distance above (Equation \ref{eq:hausdorff-dist}), the Hausdorff distance, $dist_H(\bm T_{i,K}^{(s, e)},\bm T_{i^\prime,K}^{(s^\prime, e^\prime)})$, between two agent trajectory sub-matrices, $\bm T_{i,K}^{(s, e)} = \{T_{i,1}^{(s, e)}, T_{i,2}^{(s, e)}, ..., T_{i,|K|}^{(s, e)}\}$
and $\bm T_{i^\prime,K}^{(s^\prime, e^\prime)} = \{T_{i^\prime,1}^{(s, e)}, T_{i^\prime,2}^{(s, e)}, ..., T_{i^\prime,|K|}^{(s, e)}\}$, is: 
\begin{equation*}
dist_H(\bm T_{i,K}^{(s, e)},\bm T_{i^\prime,K}^{(s^\prime, e^\prime)}) = \max{h(\bm T_{i,K}^{(s, e)},\bm T_{i^\prime,K}^{(s^\prime, e^\prime)}),h(\bm T_{i^\prime,K}^{(s^\prime, e^\prime)},\bm T_{i,K}^{(s, e)})}, \text{ where }
\end{equation*}
\begin{equation*}
h(\bm T_{i,K}^{(s, e)},\bm T_{i^\prime,K}^{(s^\prime, e^\prime)}) = \max\limits_{T_{i,k}^{(s, e)} \in \bm T_{i,K}^{(s, e)}} \min\limits_{T_{i^\prime,k}^{(s^\prime, e^\prime)} \in \bm T_{i^\prime,K}^{(s^\prime, e^\prime)}} dist(T_{i,k}^{(s, e)}, T_{i^\prime,k}^{(s^\prime, e^\prime)})
\end{equation*}
and
\begin{equation*}
h(\bm T_{i^\prime,K}^{(s^\prime, e^\prime)},\bm T_{i,K}^{(s, e)}) = \max\limits_{T_{i^\prime,k}^{(s^\prime, e^\prime)} \in \bm T_{i^\prime,K}^{(s^\prime, e^\prime)}} \min\limits_{T_{i,k}^{(s, e)} \in \bm T_{i,K}^{(s, e)}} dist(T_{i^\prime,k}^{(s^\prime, e^\prime)}, T_{i,k}^{(s, e)})
\end{equation*}

\textbf{Trajectory sub-matrix $\varepsilon$-similar-neighborhood and support.} 
The $\varepsilon$-similar-neighborhood for each sub-matrix is the set of sub-matrices within a Hausdorff distance of $\varepsilon$, and is given by:
\[
N_\varepsilon(\bm T_{i,K}^{(s,e)}) := \{\bm T_{i^\prime,K}^{(s^\prime,e^\prime)} \mid {\rm dist_H}(\bm T_{i,K}^{(s,e)}, \bm T_{i^\prime,K}^{(s^\prime,e^\prime)}) \le \varepsilon\},
\]
where $\varepsilon$ is the distance threshold, a user-selectable parameter of MA-Stat-DSM.

The support of $\bm T_{i,K}^{(s, e)}$ with respect to a subset of sub-matrices $G_m \subseteq [n]$ (where $[n]$ denotes the set of all trajectory matrices in the dataset) is:
\[
{\rm sup}_{G_m}(\bm T_{i,K}^{(s, e)}) := |\{i^\prime \in G_m \mid \exists~\bm T_{i^\prime,K}^{(s^\prime,e^\prime)} \sqsubseteq \bm T_{{i^\prime,K}}, \bm T_{i^\prime,K}^{(s^\prime,e^\prime)} \in N_\varepsilon(\bm T_{i,K}^{(s, e)})\}|
\]
This support represents the number of sub-matrices in $G_m$ that contain at least one sub-matrix with distance from sub-matrix $\bm T_{i,K}^{(s, e)} \leq \varepsilon$.

As mentioned, in this study, we consider the set of all trajectory matrices (attacks) of a particular team in a specific match, so when referring to ``all trajectory matrices in the dataset,'' in our problem setup this means ``all attacks by team $T$ in match $M$.''

\textbf{Basketball-specific example.}
Although specific plays will be interpreted for other matches in the results section, here, we describe an example of the problem setup and how MA-Stat-DSM is applied to the set of team attacks from a specific match, as depicted in Figure 2.

The Los Angeles Lakers attacks in our dataset from their 15 November 2015 match against the Detroit Pistons are shown, with the effective attacks ($G_+$) and ineffective attacks ($G_-$) shown in the top and bottom panels, respectively.
The $N = 12$ plays, which are represented as $|K|$-by-$m_i$ trajectory matrices, are shown based on the time interval from when the last passer receives the ball to when a shot is attempted, t2 to t0 (Figure 1).
There are five effective attacks, i.e., $|G_+| = 5$, and seven ineffective attacks, i.e., $|G_-| = 7$, based on our definition of effectiveness proposed in subsection 3.2.
That is, each of the top five matrices is labelled effective (+1) and each of the bottom seven matrices has an ineffective (-1) label.
Each of the twelve attacks represents an agent trajectory matrix, $\bm T_{i,K}$, which consists of the contemporaneous trajectories ($T_{i,k}$) of each of the $|K| = 5$ agents considered, i.e., $K = \{$ball, (Laker's) shooter, (Laker's) last passer, (Piston's) shooter defender, (Piston's) last passer defender\}.
The number of rows in the $i$-th agent trajectory matrix corresponds to the number of agents, $|K| = 5$, and the number of columns in each agent trajectory matrix corresponds to the length of the trajectories, $m_i$  (number of point coordinates).
As mentioned, all agent trajectories within the same trajectory matrix have the same length ($m_{i,k}$ is the same for all $k \in K$, but the length of trajectories across attacks differs.
Specifically, due to the high frequency of the SportsVU data, it is unlikely (but not impossible) that the length (number of columns) of one play is the same as another (i.e., in general, $m_i \neq m_{i^\prime}$ for $i \neq i^\prime$).

In this problem setup, MA-Stat-DSM (described in the next section) can be applied to this set of twelve labelled agent trajectory matrices to identify the portions of the plays that discriminate between the effective and ineffective labels (the SSD sub-matrices, $\bm T^{(s,e)}_{i,K}$ of play/trajectory matrix $i$, if exists).
The discriminative sub-matrices again have $|K| = 5$ rows, and contain the SSD contemporaneous sub-trajectories, $T^{(s,e)}_{i,K}$, of each agent $k \in K$.
The SSD sub-matrices are depicted plus signs.
As can be observed in Figure 2, when applying MA-Stat-DSM (with a distance threshold $\varepsilon = 4$ and a minimum length of $L = 4$) to the 12 Laker attacks from their 15 November 2015 match against the Pistons, five agent trajectory matrices contained SSD sub-matrices.
In particular, two of the effective plays had discriminative sub-matrices and three of the ineffective plays contained discriminative sub-matrices.

SSD sub-matrices in an effective agent trajectory matrix indicate the portion of the play that was relevant in rendering the attack effective rather than ineffective.
Similarly, SSD sub-matrices in an ineffective agent trajectory matrix indicate the portion of the play that was relevant in rendering the attack ineffective rather than effective.
Note that in some cases, e.g., the second effective play from the right, the discriminative portion of the play comprised all of --- or nearly all of --- the agent trajectory matrix, i.e., the SSD sub-matrix and agent trajectory matrix are the same, indicating that the whole passage of play in this time interval was relevant in rendering this attack effective rather than ineffective.


\begin{landscape}
\begin{figure}[h]
\centering
\includegraphics[width=1.55\textwidth]{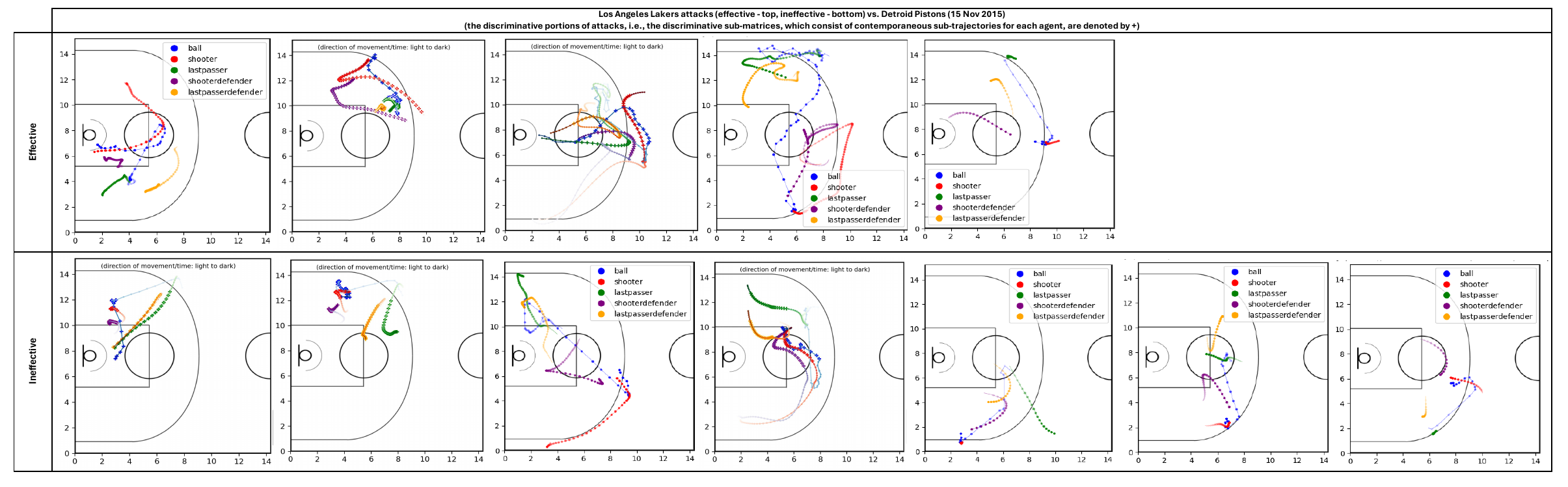}
\caption{Los Angeles Lakers (effective - top, ineffective - bottom) attacks in our dataset from their 15 November 2015 match against the Detroit Pistons. The effective and ineffective attacks, based on our definition proposed, are shown in the top and bottom panels, respectively. The SSD sub-matrices are depicted by plus signs.}
\label{fig:basketball_specific_example}
\end{figure}
\end{landscape}

\subsection{MA-Stat-DSM Algorithm}
\label{ssec:algorithm}
The flow of MA-Stat-DSM (Algorithm \ref{alg:algorithm}) pseudocode is relatively similar to Stat-DSM.
The main changes are that Euclidean distance is replaced by Hausdorff distance ($dist_H$ on lines 27 and 28 of the pseudo-code), trajectory and sub-trajectory are replaced with trajectory matrix and sub-matrix, respectively, and a fixed set of $K$ agents is considered.
Figure 3 depicts the main steps of the MA-Stat-DSM algorithm.

\begin{figure}
\centering
\includegraphics[width=0.75\textwidth]{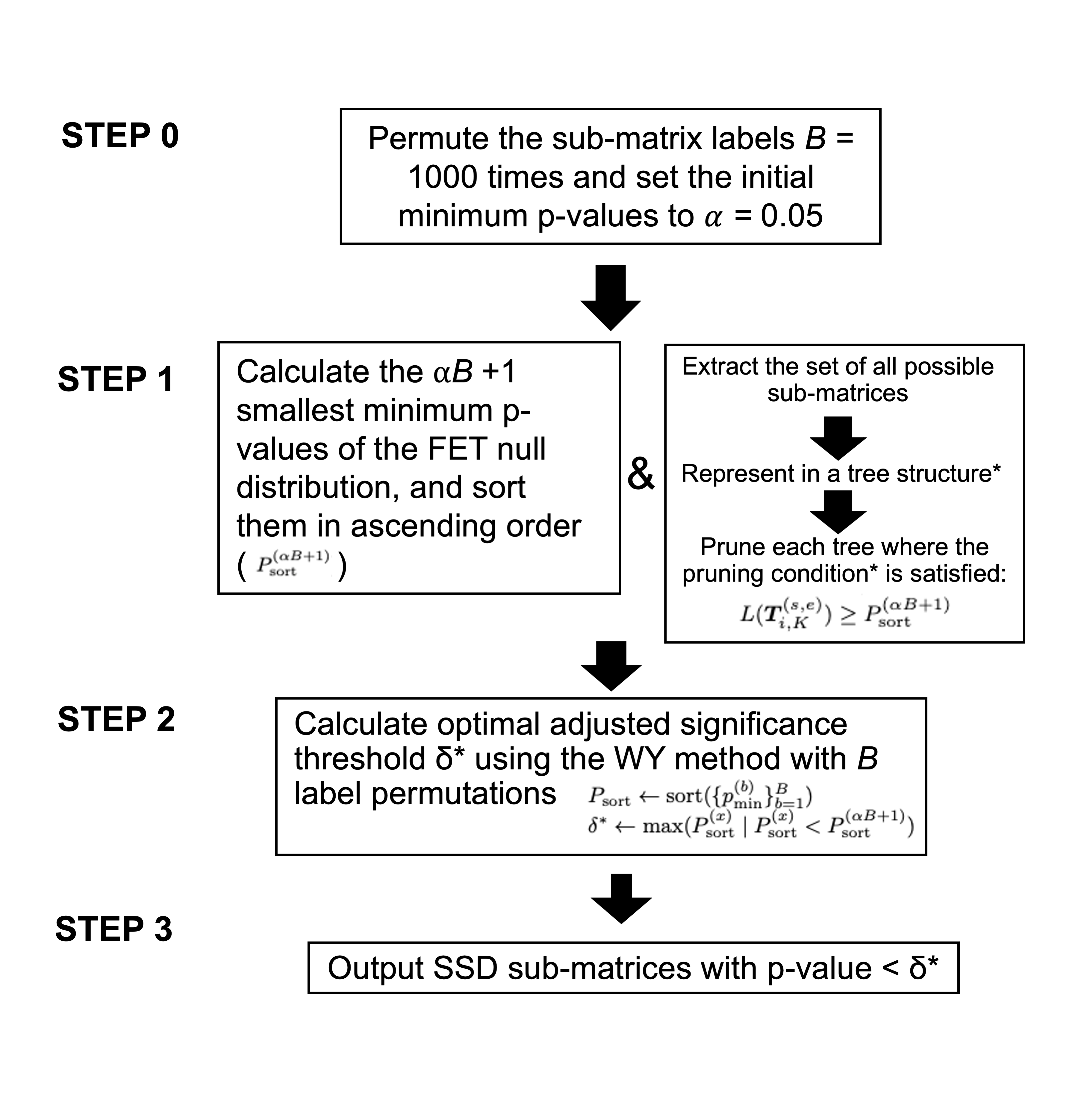}
\caption{Main steps of the MA-Stat-DSM algorithm. *See Figures 3 and 4 in \cite{le2020stat} for more details of the tree representation and pruning properties of Stat-DSM, which also apply to MA-Stat-DSM but by replacing trajectory and sub-trajectory with trajectory matrix and sub-matrix, respectively.
}
    \label{fig:mastatdsm-flow}
\end{figure}

The Python code for MA-Stat-DSM is available on GitHub (see the Appendix for the URL).

{\SetAlgoNoLine
\begin{algorithm}
\SetKwProg{Function}{function}{}{}
\SetKwProg{Procedure}{procedure}{}{}

\let\oldnl\nl
\newcommand{\nonl}{\renewcommand{\nl}{\let\nl\oldnl}}

\DontPrintSemicolon 
\KwIn{Set of $K$ agents (fixed), Trajectory matrix dataset $\bm D = \{\bm T, \bm g\}$, distance threshold $\varepsilon$, minimum length $L$, number of permutations $B$, and significance level $\alpha$.}
\KwOut{\ Statistically discriminative sub-matrices}

\Procedure{Main \rm()}{
	\nonl // Initialization\;
	\For{$b \leftarrow 1$ {\rm to} $B$}{ \label{alg:start_initialization}
		$\bm D^{(b)} \leftarrow \{\bm T,  {\rm permute}($\textbf{g}$)\}$\;
		$p_{\rm min}^{(b)} \leftarrow \alpha$\; 
	}\label{alg:end_initialization}
	\nonl // Extract sub-matrices and estimate the null distribution\;	
	\For{{\rm each} $\bm T_{i,K} \in \bm T$}{ \label{alg:start_step1}
		\For{{\rm each length-$L$} {\rm sub-matrix} $\bm T_{i,K}^{(s,e)} \sqsubseteq \bm T_{i,K}$}{
			Compute $N_\varepsilon(\bm T_{i,K}^{(s,e)})$\; \label{calculate_neighbor}
			\textit{ProcessNext {\rm (}$\bm T_{i,K}^{(s,e)}$, $N_{\varepsilon}(\bm T_{i,K}^{(s,e)})${\rm )}} 
		}
	}\label{alg:end_step1}

	\nonl // Calculate the adjusted significance level $\delta^{\ast}$\;
	$P_{\rm sort} \leftarrow {\rm sort}(\{p_{\rm min}^{(b)}\}_{b=1}^B)$\; \label{alg:start_calibrate}
	$\delta^{\ast} \leftarrow \max(P_{\rm sort}^{(x)}\ |\ P_{\rm sort}^{(x)} < P_{\rm sort}^{(\alpha B+1)})$ \; \label{alg:end_calibrate}
	
	\nonl // Statistically discriminative sub-matrices\;
	Output the sub-matrices with \textit{p}-values $ < \delta^{\ast}$ \label{alg:enumerate_discriminative}
}
\BlankLine
\Function{ProcessNext {\rm (}$\bm T_{i,K}^{(s,e)}$, $N_{\varepsilon}(\bm T_{i,K}^{(s,e)})${\rm )}}{
	$P_{\rm sort} \leftarrow {\rm sort}(\{p_{\rm min}^{(b)}\}_{b=1}^B)$\; \label{alg:sort_list_p}
	Compute $L(\bm T_{i,K}^{(s,e)})$\; \label{alg:compute_lower_bound}

	\If{$L(\bm T_{i,K}^{(s,e)}) \ge P_{\rm sort}^{(\alpha B+1)}$}{ \label{alg:start_pruning}
		\Return\;
	}\label{alg:end_pruning}

	\For{$b \leftarrow 1$ {\rm to} $B$}{ \label{alg:start_update_min_p}
		\If{$L(\bm T_{i,K}^{(s,e)}) < p_{\rm min}^{(b)}$}{
			$ p_{\rm min}^{(b)} \leftarrow \min\{\ p_{\rm min}^{(b)}, p^{(b)}(\bm T_{i,K}^{(s,e)})\ \}$
		}
	}  \label{alg:end_update_min_p}

	\For{{\rm each} $\bm T_{{i^\prime,K}}^{(s^\prime, e^\prime)} \in N_{\varepsilon}(\bm T_{i,K}^{(s,e)})$}{ \label{alg:start_continue_extraction}
		$d \leftarrow {\rm dist_H}(\bm T_{{i^\prime,K}}^{(s^\prime, e^\prime + 1)}, \bm T_{i,K}^{(s, e + 1)})$\;
		\If{$d \leq \varepsilon$}{
			Add $\bm T_{{i^\prime,K}}^{(s^\prime, e^\prime + 1)}$ into $N_{\varepsilon}(\bm T_{i,K}^{(s, e + 1)})$\;
		}
	} 
	\textit{ProcessNext{\rm (}$\bm T_{i,K}^{(s, e + 1)}$, $N_{\varepsilon}(\bm T_{i,K}^{(s, e + 1)})${\rm )}} \label{alg:end_continue_extraction}
}

\caption{Multi-Agent Statistically Discriminative Sub-trajectory Mining (MA-Stat-DSM)}
\label{alg:algorithm}
\end{algorithm}}

\section{Results}
\label{sec:results}

In this section, we first outline the experimental setup in terms of the parameters selected for MA-Stat-DSM in subsection \ref{ssec:experimental-setup}.
%
%
Then, we provide examples of effective and ineffective discriminative SSD sub-matrix results that were obtained for matches involving two of the top teams from the 2015/2016 NBA competition (Cleveland and Golden State) and the bottom teams from the Eastern and Western Conferences (the Philadelphia 76ers and Los Angeles Lakers) and interpret the results from a practical perspective.

\subsection{Experimental Setup}
\label{ssec:experimental-setup}
Although there was limited \textit{a-priori} knowledge as to what the appropriate parameter setting for MA-Stat-DSM should be for the basketball dataset we are using, the paper in which Stat-DSM was proposed \cite{le2020stat} --- which applied the proposed method to datasets consisting of vehicle and hurricane trajectories --- was used as a starting point. 
While these datasets involved greater distances than those on a basketball court, to balance this, the data was also much lower frequency than the NBA trajectory data. 
Thus, as a starting point, we consider the range of parameters specified for Stat-DSM by \cite{le2020stat}  (Table \ref{tab:parameter-setting}).
%
%
%
Ultimately, \cite{le2020stat} selected a significance level, $\alpha$, of 0.05 and the number of permutations, $B$, to be 1,000, and we also selected these values in the present study.
%

\begin{table}[!ht]
\caption{Parameter setting for Stat-DSM from \cite{le2020stat}.}
\label{tab:parameter-setting}
\begin{tabular}{llll}
Minimum length $L$ & \multicolumn{3}{c}{5, 8, 10}\\
Distance threshold $\varepsilon$ &  \multicolumn{3}{c}{1.5, 4, 20}\\
Number of permutations $B$ & \multicolumn{3}{c}{1000}\\
Significance level $\alpha$ & \multicolumn{3}{c}{0.05}\\
\end{tabular}
\end{table}

%
%
%

%

The MA-Stat-DSM parameter values we used are shown in Table \ref{tab:parameter-setting-ma}.
MA-Stat-DSM was applied to the trajectory data, with a statistical significance level of 0.05, $B = 1000$, distance thresholds of 1.5 and 4, and minimum lengths of 5, 8 and 10.
The data preprocessing parameters used were $|K|$ = 5 agents and a time interval from t2 to t0 (Figure 1). 
%
%

Table 3 shows the number of SSD matrices within attacks (and the number of distinct matches containing those attacks) obtained with the parameter settings in Table 2.
It can be observed that, with a distance threshold of 1.5, only very few SSD sub-matrices were obtained by MA-Stat-DSM compared to when a distance threshold of 4 was used.
This suggests that at the match level, a distance threshold of 4 is more appropriate to ensure an adequate number of SSD sub-matrices within attacks can be obtained.
Comparing the ``No. of distinct matches'' with the SSD attacks by the ``No. matches'' column in Table 3 shows that most of the teams' matches within the season contained some SSD attack(s).
%
%
While the analysis in Table 3 is useful for selecting appropriate parameters, note that it does not show the number of attacks with SSD sub-matrices within each match.
It should also be noted that there is overlap in the attack and match counts in Table 3, e.g., sub-matrices and attacks/matches with a minimum length of 5 can also be obtained with a minimum length of 8 and 10.

The experiments using the MA-Stat-DSM algorithm were run on an Intel(R) Xeon(R) CPU E5-2697 v2 @ 2.70GHz Linux CentOS server machine (Linux version 3.10.0), running at CPU 3.2 GHz, 
using 128 GB of RAM.
The most influential factor affecting the run time of the MA-Stat-DSM algorithm was the distance threshold parameter, taking nearly five times as long on average (per MA-Stat-DSM iteration) to run the algorithm with a distance threshold of 4 compared to a distance threshold of 1.5.
A distance threshold of 20 was not feasible on our dataset because of the computational complexity, which was found to be prohibitive when running MA-Stat-DSM on a particular team's attacks in a specific match.

\begin{table}[!ht]
\caption{MA-Stat-DSM parameters (top) and data preprocessing parameters (bottom)}
\label{tab:parameter-setting-ma}
\begin{tabular}{rrrr}
Minimum length $L$ & \multicolumn{3}{c}{5, 8, 10}\\
Distance threshold $\varepsilon$ &  \multicolumn{3}{c}{1.5, 4}\\
Number of permutations $B$ & \multicolumn{3}{c}{1000}\\
Significance level $\alpha$ & \multicolumn{3}{c}{0.05}\\ \hline
Set of agents $K$ & \multicolumn{3}{c}{\{ball, shooter, last passer, shooter defender, last passer defender\}}\\
Time interval & \multicolumn{3}{c}{t2 to t0 (as per Figure 1)}\\
\end{tabular}
\end{table}




\begin{sidewaystable}
\centering
\caption{Number of matches containing attacks with statistically significantly discriminative (SSD) sub-matrices, and the number of attacks containing SSD sub-matrices, for $\varepsilon = 1.5$ and $\varepsilon = 4$, and $L = 5, L = 8, L = 10$ for the two top and two bottom teams' matches in the 2015/16 season.}
\resizebox{\linewidth}{!}{%
\begin{tabular}{>{\centering\hspace{0pt}}m{0.04\linewidth}>{\centering\hspace{0pt}}m{0.044\linewidth}>{\centering\hspace{0pt}}m{0.052\linewidth}>{\centering\hspace{0pt}}m{0.048\linewidth}>{\centering\hspace{0pt}}m{0.096\linewidth}>{\centering\hspace{0pt}}m{0.088\linewidth}>{\centering\hspace{0pt}}m{0.088\linewidth}>{\centering\hspace{0pt}}m{0.09\linewidth}>{\centering\hspace{0pt}}m{0.117\linewidth}>{\centering\hspace{0pt}}m{0.117\linewidth}>{\centering\arraybackslash\hspace{0pt}}m{0.117\linewidth}} 
\toprule
team                                                      & No. matches                                               & Distance threshold ($\varepsilon$) & No. SSD attacks & No. of distinct matches containing attacks with SSD sub-matrices & No. of attacks with SSD sub-matrices ($L=5$) & No. of~attacks with SSD sub-matrices ($L=8$) & No. of attacks with SSD sub-matrices ($L=10$) & No. of distinct matches containing attacks with SSD sub-matrices ($L=5$) & No. of distinct matches containing attacks with SSD sub-matrices ($L=8$) & No. of distinct matches containing attacks with SSD sub-matrices ($L=10$)  \\ 
\midrule
\multirow{2}{0.04\linewidth}{\hspace{0pt}\Centering{}GSW} & \multirow{2}{0.044\linewidth}{\hspace{0pt}\Centering{}40} & 4                  & 625             & 40                                                               & 242                                               & 198                                               & 185                                                & 33                                                                            & 30                                                                            & 31                                                                              \\ 
\cline{3-11}
                                                          &                                                           & 1.5                & 4               & 4                                                                & 2                                                 & 1                                                 & 1                                                  & 2                                                                             & 1                                                                             & 1                                                                               \\ 
\midrule
\multirow{2}{0.04\linewidth}{\hspace{0pt}\Centering{}PHI} & \multirow{2}{0.044\linewidth}{\hspace{0pt}\Centering{}40} & 4                  & 815             & 35                                                               & 306                                               & 270                                               & 239                                                & 34                                                                            & 33                                                                            & 31                                                                              \\ 
\cline{3-11}
                                                          &                                                           & 1.5                & 4               & 4                                                                & 2                                                 & 1                                                 & 1                                                  & 2                                                                             & 1                                                                             & 1                                                                               \\ 
\midrule
\multirow{2}{0.04\linewidth}{\hspace{0pt}\Centering{}LAL} & \multirow{2}{0.044\linewidth}{\hspace{0pt}\Centering{}41} & 4                  & 786             & 31                                                               & 294                                               & 255                                               & 237                                                & 31                                                                            & 29                                                                            & 28                                                                              \\ 
\cline{3-11}
                                                          &                                                           & 1.5                & 10              & 4                                                                & 6                                                 & 2                                                 & 2                                                  & 4                                                                             & 1                                                                             & 1                                                                               \\ 
\midrule
\multirow{2}{0.04\linewidth}{\hspace{0pt}\Centering{}CLE} & \multirow{2}{0.044\linewidth}{\hspace{0pt}\Centering{}36} & 4                  & 692             & 30                                                               & 244                                               & 229                                               & 219                                                & 28                                                                            & 28                                                                            & 27                                                                              \\ 
\cline{3-11}
                                                          &                                                           & 1.5                & 2               & 2                                                                & 2                                                 & 0                                                 & 0                                                  & 2                                                                             & 0                                                                             & 0                                                                               \\
\bottomrule
\end{tabular}
}
\end{sidewaystable}

\subsection{Visualization and interpretation of SSD sub-matrix examples}
\label{ssec:result-intepretation}
In this subsection, we provide some SSD sub-matrix result examples that were obtained when applying MA-Stat-DSM to team attacks in selected matches (listed in Table 4) involving the two top teams and two bottom teams in the 2015/16 NBA season.
Table 3 shows that a minimum length parameter of 5 obtained more SSD attacks within matches than $L$ = 8 or $L$ = 10.
Furthermore, as mentioned previously, a distance threshold of 4 is preferred when applying MA-Stat-DSM at the match level.
The remainder of the parameters for the following results are $B = 1000$, $\alpha = 0.05$, $K = \{$ball, shooter, shooter defender, last passer, last passer defender$\}$, and with the time interval of t2 to t0 (Figure 1).


\begin{table}
\centering
\caption{Number of attacking plays in the match and the number of those that contained SSD sub-matrices for selected matches involving the top- and bottom-performing teams in the 2015/16 NBA season (with $\varepsilon=4$, $L = 5$).}
\resizebox{\linewidth}{!}{%
\begin{tabular}{>{\RaggedLeft\hspace{0pt}}m{0.079\linewidth}>{\hspace{0pt}}m{0.106\linewidth}>{\hspace{0pt}}m{0.102\linewidth}>{\RaggedLeft\hspace{0pt}}m{0.106\linewidth}>{\RaggedLeft\hspace{0pt}}m{0.104\linewidth}>{\RaggedLeft\hspace{0pt}}m{0.21\linewidth}>{\RaggedLeft\hspace{0pt}}m{0.21\linewidth}} 
\toprule
\multicolumn{1}{>{\hspace{0pt}}m{0.079\linewidth}}{\textbf{Match Date}} & \textbf{Home Team (HT)} & \textbf{Away Team (AT)} & \multicolumn{1}{>{\hspace{0pt}}m{0.106\linewidth}}{\textbf{No. of HT attacks}} & \multicolumn{1}{>{\hspace{0pt}}m{0.104\linewidth}}{\textbf{No. of AT attacks}} & \multicolumn{1}{>{\hspace{0pt}}m{0.21\linewidth}}{\textbf{No. of HT attacks with SSD sub-matrices}} & \multicolumn{1}{>{\hspace{0pt}}m{0.21\linewidth}}{\textbf{No. of AT attacks with SSD sub-matrices}}  \\ 
\midrule
10-Jan-16                                                               & PHI                     & CLE                     & 17                                                                             & 20                                                                             & 4                                                                                                   & 2                                                                                                    \\ 
\midrule
05-Jan-16                                                               & LAL                     & GSW                     & 20                                                                             & 32                                                                             & 14                                                                                                  & 17                                                                                                   \\ 
\midrule
25-Dec-15                                                               & GSW                     & CLE                     & 20                                                                             & 28                                                                             & 1                                                                                                   & 23                                                                                                   \\ 
\midrule
20-Dec-15                                                               & CLE                     & PHI                     & 20                                                                             & 24                                                                             & 2                                                                                                   & 8                                                                                                    \\ 
\midrule
01-Dec-15                                                               & PHI                     & LAL                     & 23                                                                             & 19                                                                             & 11                                                                                                  & 3                                                                                                    \\
\bottomrule
\end{tabular}
}
\end{table}
The discriminative portions of the attacks (the SSD sub-matrices) are denoted with plus signs and the remainder of the trajectories are also shown as per the time interval t2 to t0, which is the time interval from when the last passer received the ball.
The movement of the agents on the court is indicated by the colour's progression from light to dark.
Recall that the SSD sub-matrix consists of the sub-trajectories of each of the agents at the same timestamps (i.e., the columns are temporally aligned), and it represents the portion of the attack that discriminates between the effective and ineffective labeled attacks (represented by trajectory matrices) for a particular team in a specific particular match.

An SSD sub-matrix from the 5 January 2016 match between the Golden State Warriors and the Los Angeles Lakers, which was obtained with MA-Stat-DSM applied to Golden State's attacks in this match, is displayed in Figure 4 (As shown in Table 4, 17 Golden State attacks containing SSD sub-matrices were obtained in this particular match with the above-mentioned parameters, of which Figure 4 is one).
In this Golden State attack against the Lakers, the discriminative sub-trajectory of the Golden State shooter is much more spread out relative (indicating rapid movement) to the slower responding movement of the shooter defender, which enabled the shooter to reach the edge of the circle to subsequently be in a position to make a 3-point shot, which led to the attack being effective (the shot was also ultimately successful).

%
%

%

\begin{figure}[h]
\centering
\includegraphics[width=1\textwidth]{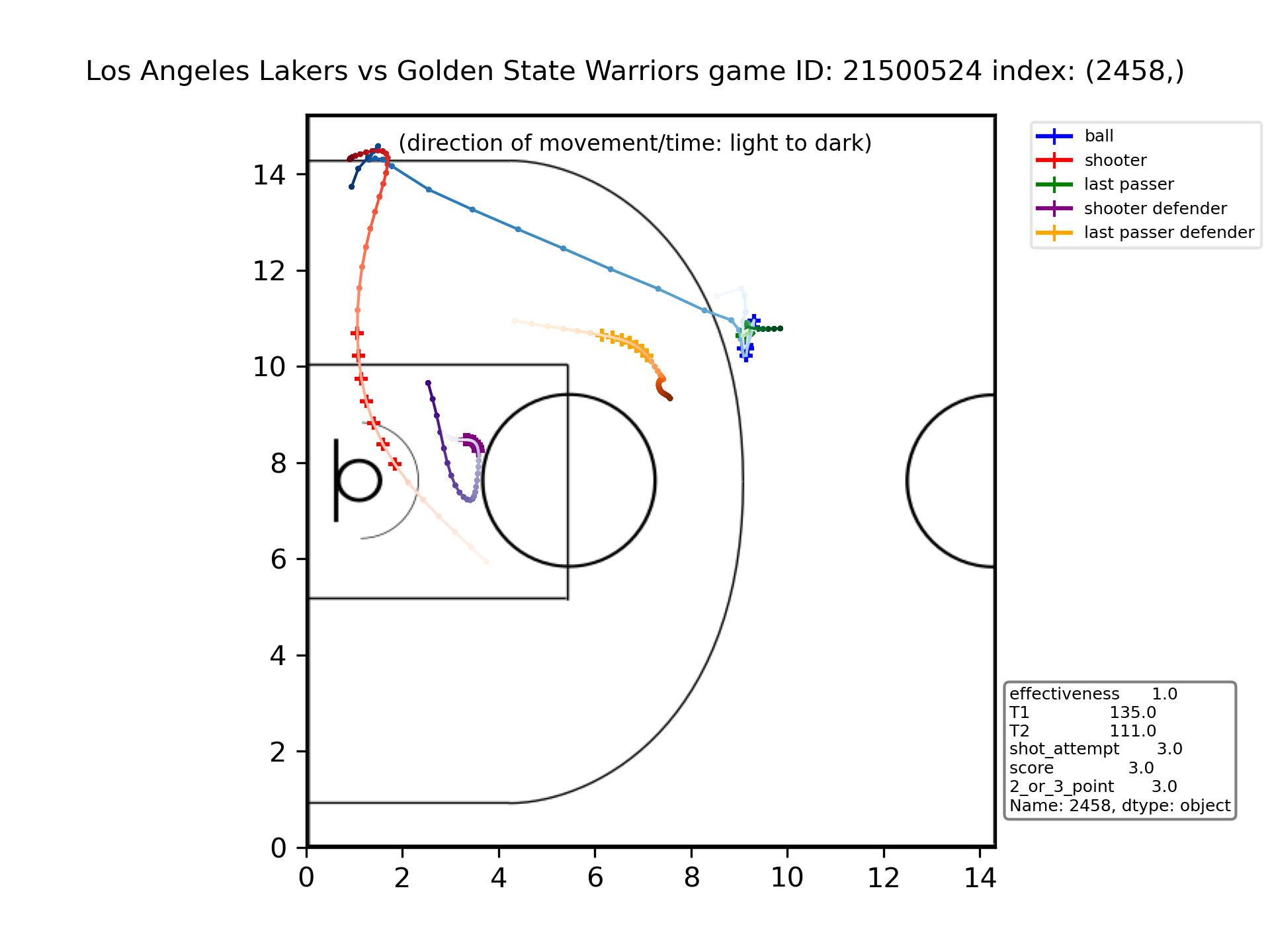}
\caption{An effective Golden State attack with an SSD sub-matrix result from the 5 January 2016 match between the Golden State Warriors and Los Angeles Lakers. The agent sub-trajectories constituting the sub-matrix are denoted by plus signs.}
\label{fig:disc_pts_1610612744_21500524_5_4_0__3_2458}
\end{figure}

An SSD sub-matrix in a Lakers attack, obtained by applying MA-Stat-DSM to the same 5 January 2016 match between the Golden State Warriors and the Los Angeles Lakers, with the parameters mentioned, is shown in Figure 5 (As shown in Table \ref{tab:parameter-setting-ma}, a 14 Lakers attacks containing SSD sub-matrices were obtained in this particular match, of which Figure 5 is one).
In this play, the SSD portion of the play shows the movement of the Lakers shooter with the ball. The speed of the movement of the drive forward by the Lakers shooter slightly outpacing the defensive movement in the same direction of the shooter defender, thereby rendering the attack effective (although a shot attempt was made, the shot itself was ultimately unsuccessful).


\begin{figure}[h]
\centering
\includegraphics[width=1\textwidth]{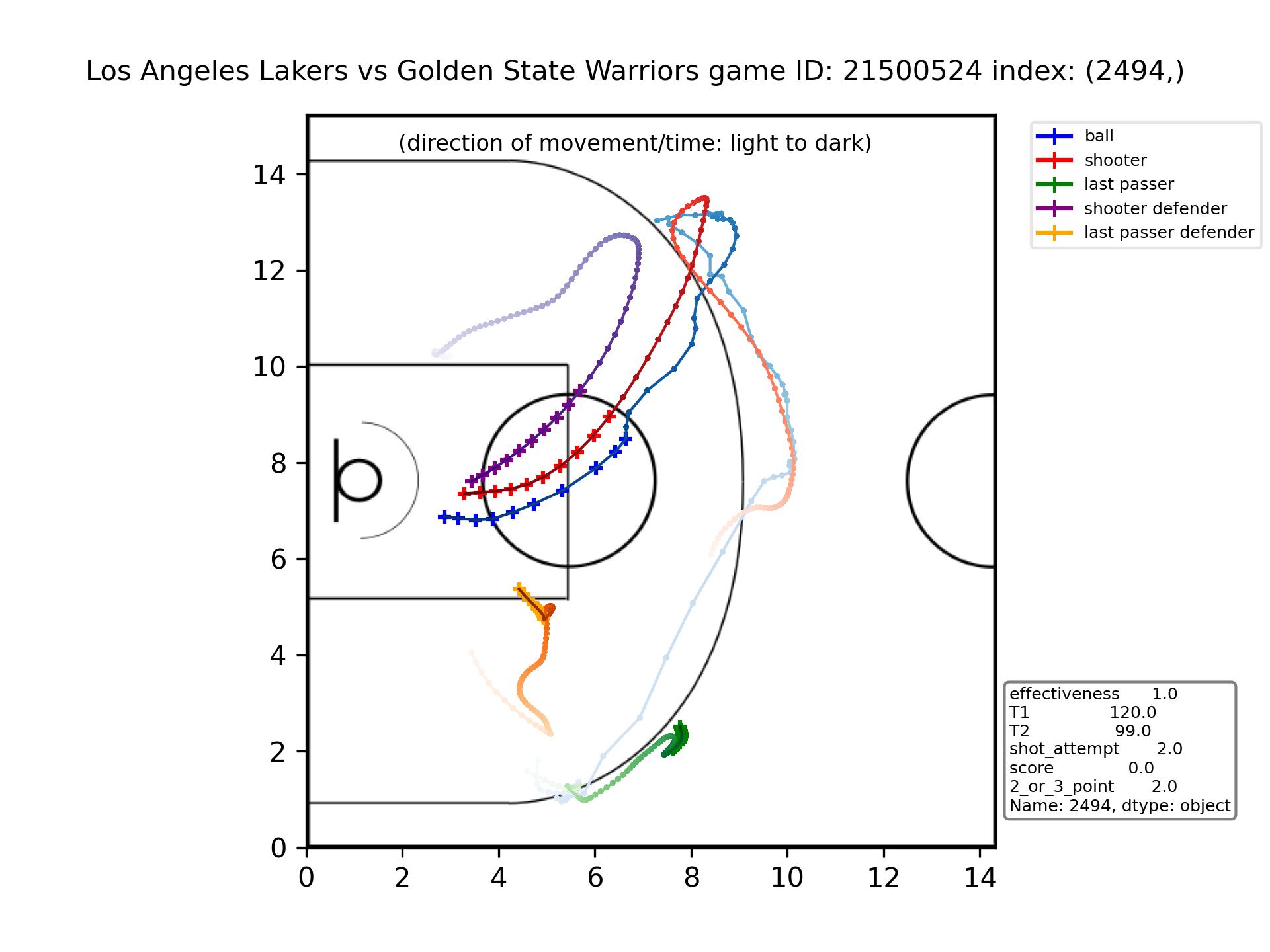}
\caption{An effective Lakers attack with an SSD sub-matrix from the 5 January 2016 match between the Golden State Warriors and Los Angeles Lakers.}
\label{fig:disc_pts_1610612747_21500524_5_4_0__10_2494}
\end{figure}

In both of the preceding examples, player movement --- off-the-ball movement of the shooter in the first case, and on-the-ball movement in the second example --- appeared to be the primary factor that resulted in the attacks being effective.
The movement of the ball through passing is another key factor that determines whether a play is ultimately effective or ineffective.
Figure 6 shows an ineffective Golden State attack, from the same 5 January 2016 against the Lakers, in which the SSD sub-matrix covers most of the agent movements within the T2 interval.
In this play, the pass made by the shooter while moving in a backward direction while being pursued by the shooter defender results in a pass that appears to be covered by the last passer defender, rendering the attack ineffective (a shot was attempted but was unsuccessful).


\begin{figure}[h]
\centering
\includegraphics[width=1\textwidth]{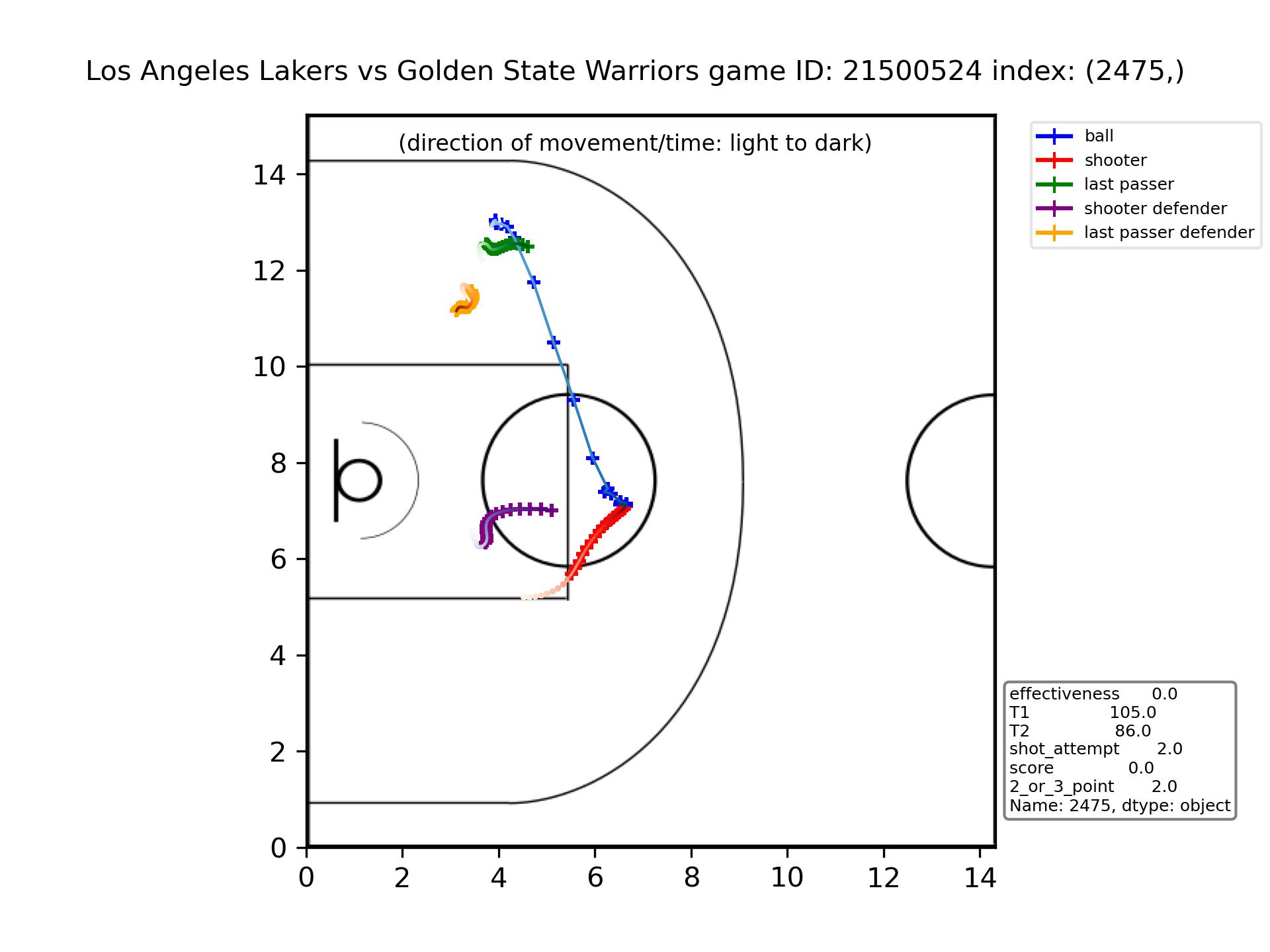}
\caption{An ineffective Golden State attack SSD sub-matrix from the 5 January 2016 match between the Golden State Warriors and Los Angeles Lakers.}
\label{fig:disc_pts_1610612744_21500524_5_4_0__11_2475}
\end{figure}

Figure 7 shows an ineffective Lakers attack SSD sub-matrix from the same match against Golden State in which the pass from open space from the last passer to the shooter does not constitute any part of the discriminative portion of the attack.
The position of the shooter at the time they receive the pass appears to be unfavourable, however, and is covered by the shooter defender.
At the same time, the last passer was making a rapid run off the ball in the circle, and it may have been a better option to pass the ball back to them, e.g., to make a lay-up, rather than making a 3-point shot attempt outside the circle under defensive pressure, which was ultimately unsuccessful.

\begin{figure}[h]
\centering
\includegraphics[width=1\textwidth]{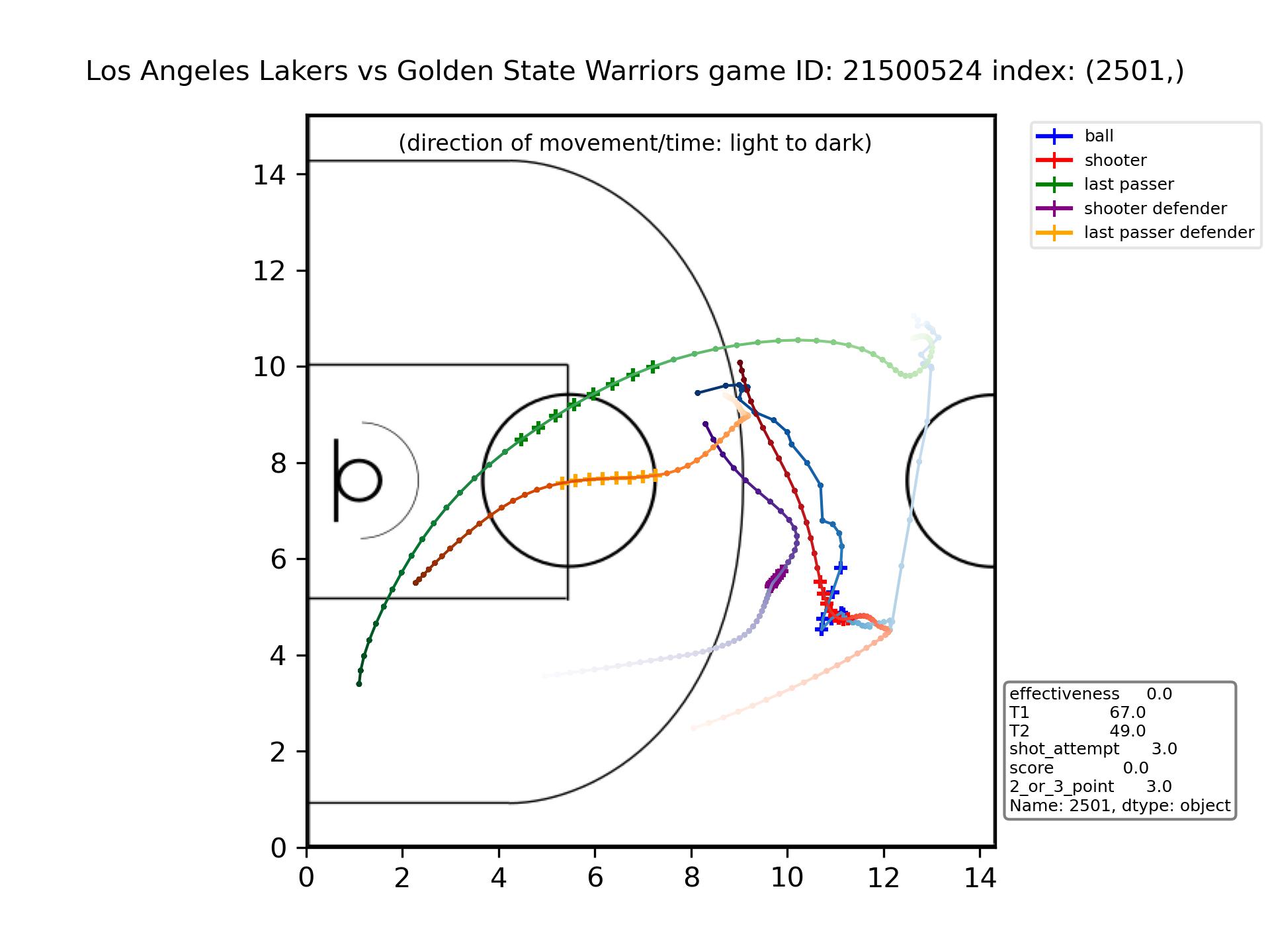}
\caption{An ineffective Lakers attack SSD sub-matrix from the 5 January 2016 match between the Golden State Warriors and Los Angeles Lakers.}
\label{fig:disc_pts_1610612747_21500524_5_4_0__13_2501}
\end{figure}


In the discriminative portion of the Lakers attack shown in Figure 8, from a match against the Philadelphia 76ers on 1 December 2015, the Laker's last passer makes a pass with roughly a 45-degree angle to the free-throw line while the shooter simultaneously makes a run perpendicular to the free-throw lane line to receive the ball from the last passer, whose pass managed to avoid the last passer defender despite the last passer defender closely tracking the last passer.
During this time, the discriminative portion of the shooter defender was relatively static, and although the play was effective, the shot attempt was unsuccessful.

\begin{figure}[h]
\centering
\includegraphics[width=1\textwidth]{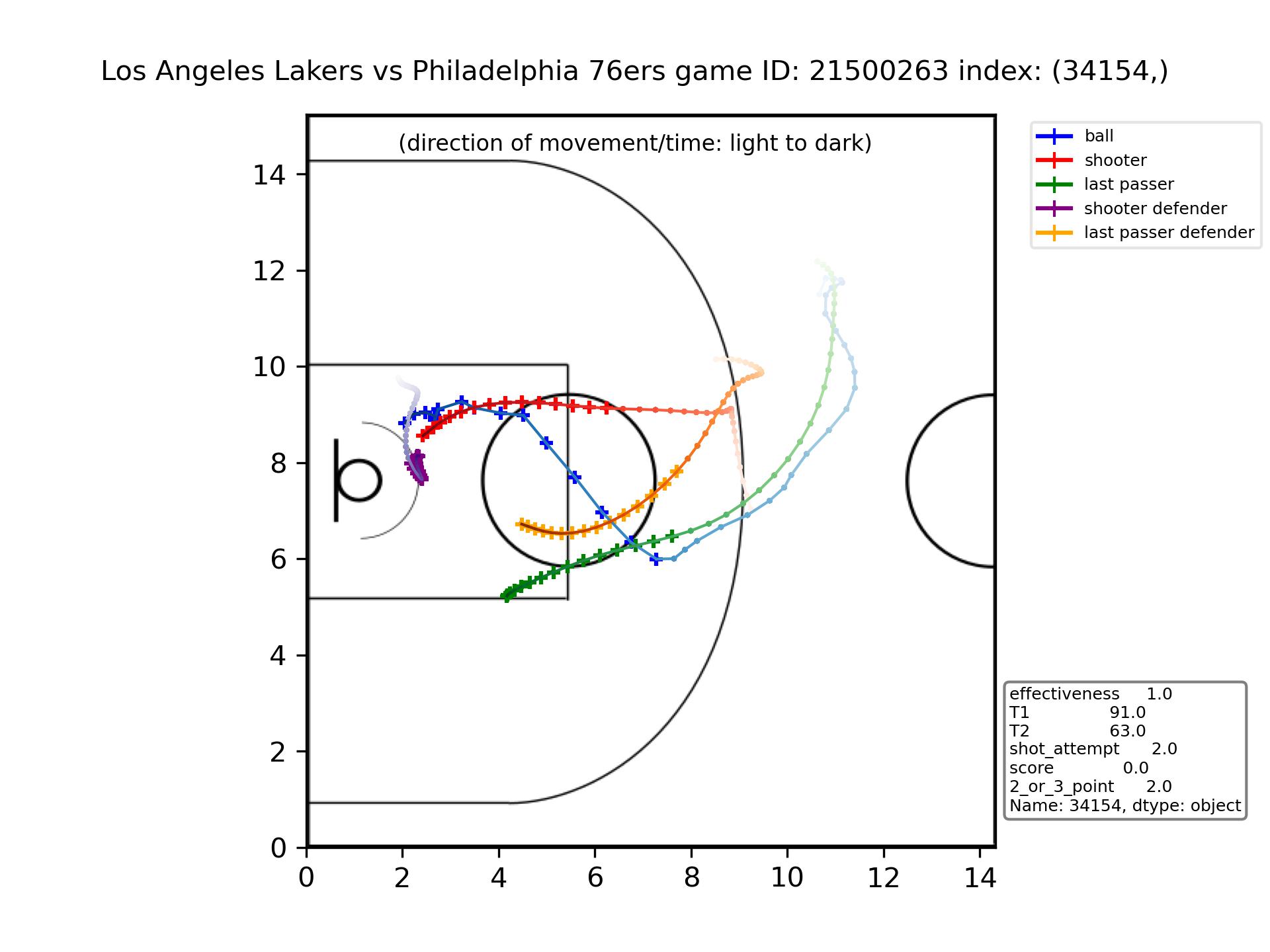}
\caption{An effective Lakers attack SSD sub-matrix from the 1 December 2015 match between the Philadelphia 76ers and Los Angeles Lakers.}
\label{fig:disc_pts_1610612747_21500263_5_4_0__1_34154}
\end{figure}

The last passer's pass to the shooter does not comprise any part of the discriminative portion of the Cleveland attack shown in Figure 9.
The discriminative portion that rendered the play ineffective was the more rapid speed of the movement of the shooter defender towards the shooter (relative to the speed of the shooter's movement), who had moved backwards to the edge of the circle after receiving the pass to attempt an unsuccessful 3 pointer.

\begin{figure}[h]
\centering
\includegraphics[width=1\textwidth]{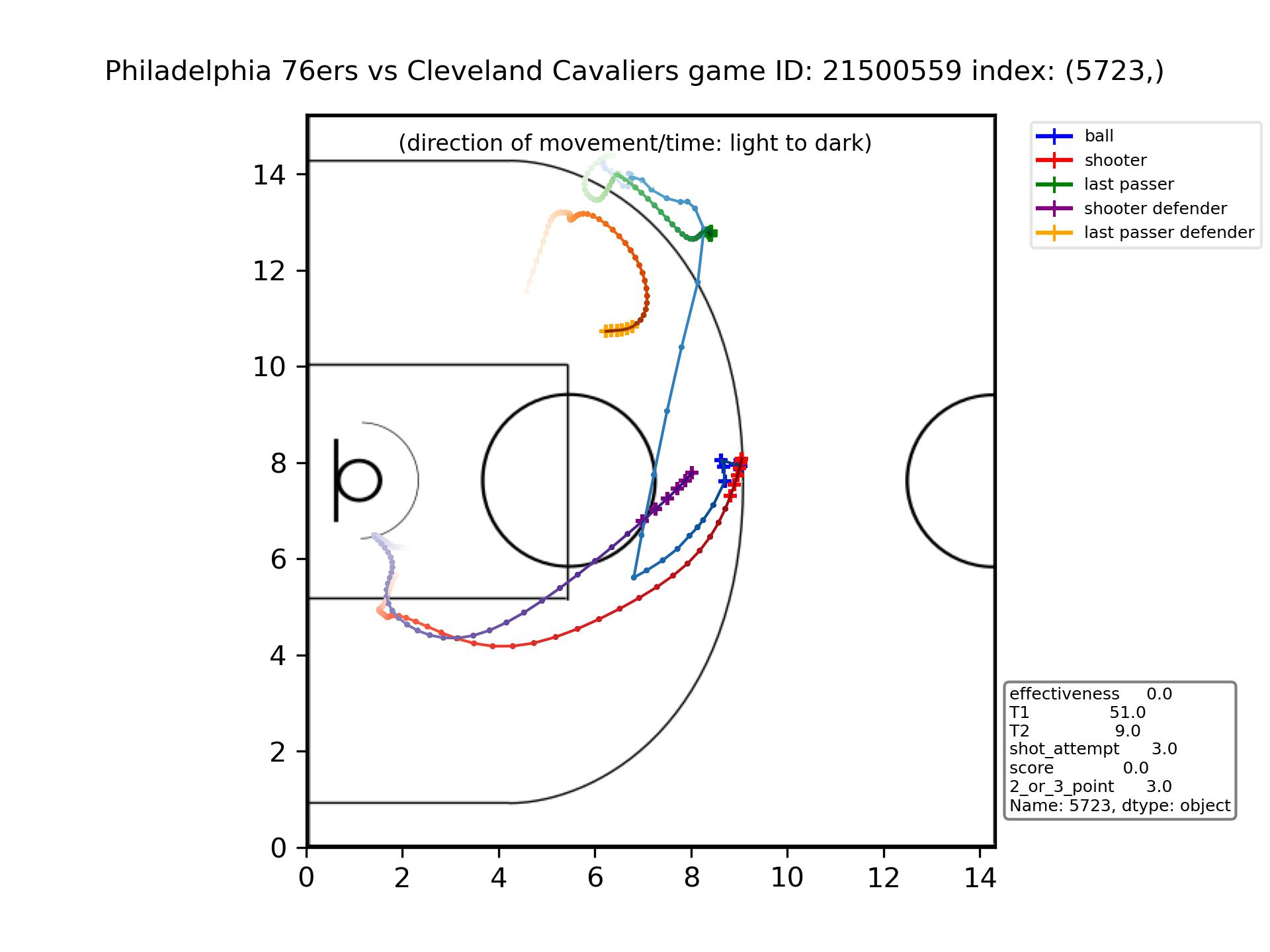}
\caption{An ineffective Cleveland attack SSD sub-matrix from the 10 January 2016 match between the Cleveland Cavaliers and Philadelphia 76ers.}
\label{fig:disc_pts_1610612739_21500559_5_4_0__0_5723}
\end{figure}

Similarly, the pass did not form any part of the discriminative sub-matrix in the ineffective 76ers attack shown in Figure 10.
The most interesting sub-trajectories in this attack seem to be those of the last passer and last passer defender, the latter tracking the movement of the last passer on the inside into the free-throw lane, perhaps meaning that the shooter could not get a pass away to them and therefore instead attempted a shot that missed from just outside the in-the-paint area.

\begin{figure}[h]
\centering
\includegraphics[width=1\textwidth]{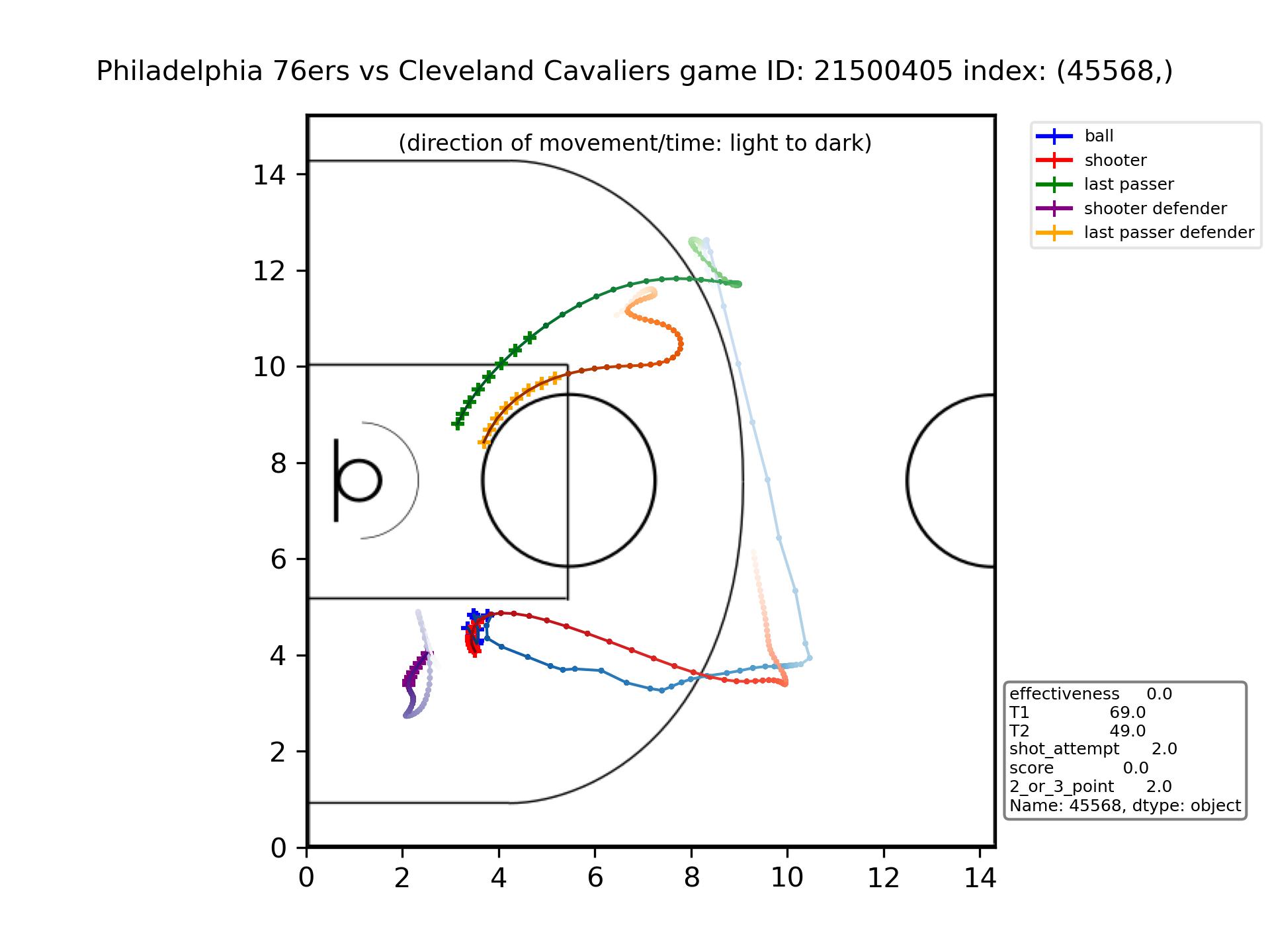}
\caption{An ineffective 76ers attack SSD sub-matrix from the 20 December 2015 match between the Cleveland Cavaliers and Philadelphia 76ers.}
\label{fig:disc_pts_1610612739_21500405_5_4_0__6_45568}
\end{figure}

Similar to the attack in Figure 6, sometimes the discriminative sub-matrix formed all --- or nearly all --- of the trajectory matrix during the time interval from t2 to t0.
In Figure 11, the trajectory sub-matrix and matrix were the same, indicating that the whole attack was relevant in discriminating between, in this case, the effective and ineffective attacks of Cleveland in their 25 December match against Golden State (this match was actually played on the 26th of December local time, but was showing in our dataset as the 25th of December).
This particular attack ended in a successful 3-point shot by the Cleveland shooter, although the distance between the shooter and shooter defender by the time the shot was made meant the play was labelled as ineffective.

\begin{figure}[h]
\centering
\includegraphics[width=1\textwidth]{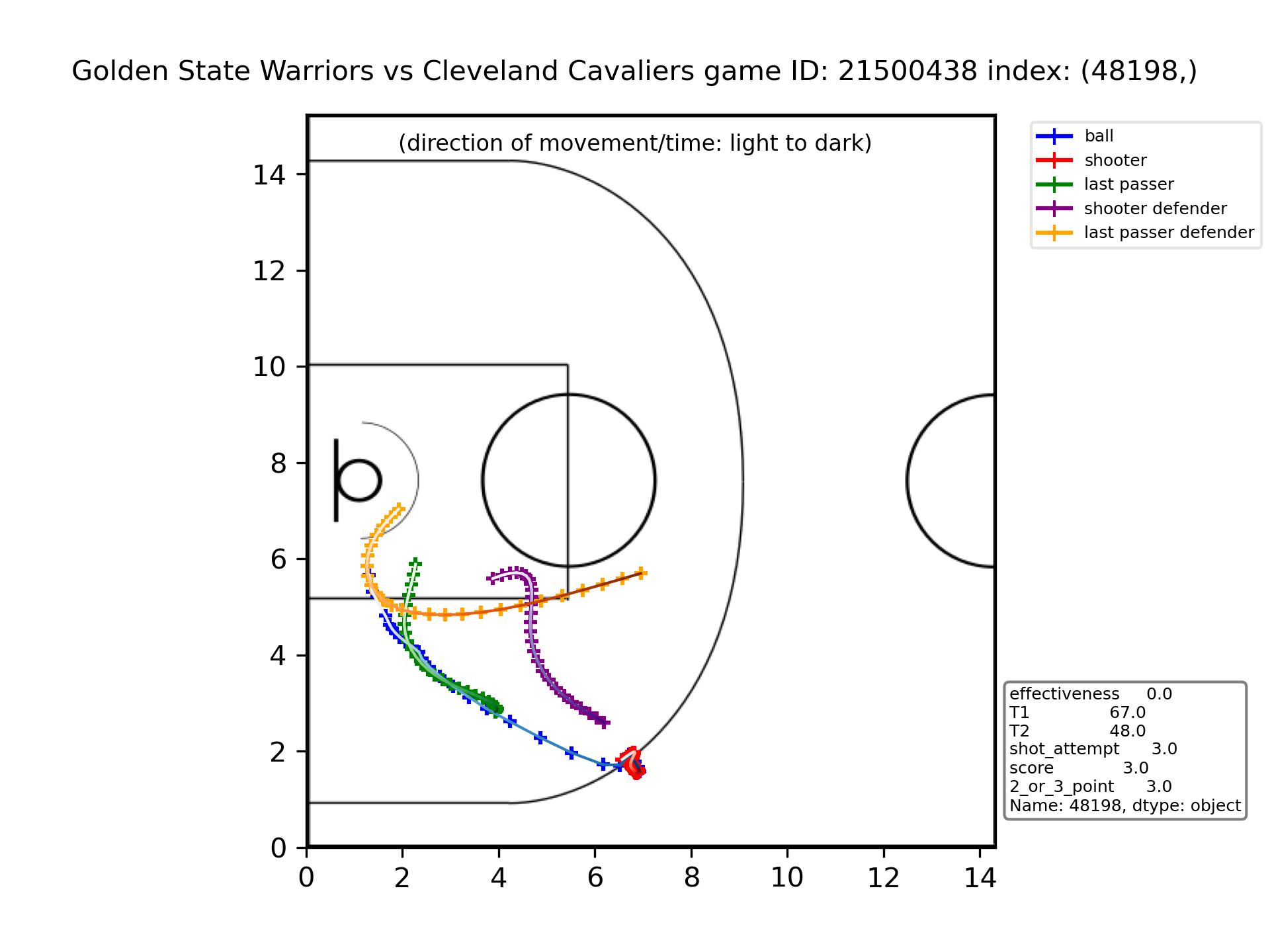}
\caption{An ineffective (but 3-point scoring) Cleveland attack SSD sub-matrix from the 25 December 2015 match between the Cleveland Cavaliers and Golden State Warriors.}
\label{fig:disc_pts_1610612739_21500438_5_4_0__17_48198}
\end{figure}

\section{Discussion}
\label{sec:discussion}
In this paper, a Multi-Agent Statistically Discriminative Sub-trajectory Mining (MA-Stat-DSM) method was proposed, which extends the Stat-DSM \cite{le2020stat} method to take trajectory matrices as input.
MA-Stat-DSM was applied to Stats Perform SportVU NBA trajectory data to identify the most relevant parts of attacking plays by identifying statistically significantly discriminative trajectory sub-matrices that discriminate between effective and ineffective attacks.
Effective attacks were defined based on the concept of wide-open shots, incorporating the position of the shooter on the court when they attempt the shot, their distance from the closest defender, and, in the case where they are attempting a three-point shot, their historical shot success probability.
While Euclidean distance was used to compute the distance between (sub)-trajectories in Stat-DSM, to compute the similarity of different trajectory (sub-)matrices, it was necessary to use an alternative distance metric, Hausdorff distance, which could compute the distance between trajectory (sub-)matrices that have different lengths (different numbers of columns).

The proposed MA-Stat-DSM method was demonstrated by applying it to attacks of a particular team in a specific match, with the labeled trajectory matrices (attacks) comprised of the trajectories of five agents (the ball and four players: the shooter, shooter defender, last passer, and last passer defender) in matches involving the two top-performing teams in the 2015/16 season (Cleveland and Golden State) and/or the bottom teams from the Eastern and Western Conferences (the 76ers and Lakers, respectively) in that season. 
%
%
The attacking plays were trimmed to consider the time interval from the time at which the last passer received the ball until the shooter made a shot attempt (or a turnover occurs such that the opposition team comes to be in possession of the ball).

The visualization of some of the SSD sub-matrices from matches involving top and bottom teams from the 2015/16 NBA season showed some cases of the sub-matrix encompassing all --- or nearly all -- of the attack trajectory matrix.
It was found that in some cases, the discriminative portion of the attack involved player movement only, i.e., the pass had already been completed and the shooter was in possession of the ball.
Alternatively, it was also found that the method was able to identify relevant off-the-ball movements by shooters (e.g., Figure 4).
In other attacks, the discriminative portion included a pass within it.
It is also possible of course, that the discriminative portion of the attack includes both a pass as well as player movement (e.g., Figure 6).

The obtained results suggested that the distance threshold is the key parameter of MA-Stat-DSM in determining how many statistically significantly discriminative (SSD) sub-matrices were obtained by the algorithm.
On the datasets to which MA-Stat-DSM was applied in the current study, namely, the attacks of a specific team in a particular match, a distance threshold of 4 was found to be more appropriate than a distance threshold of 1.5 in order to obtain an adequate number of SSD sub-matrices.
Even when the MA-Stat-DSM was iterated over all of a team's matches in the season, a distance threshold only obtained a very small number of SSD sub-matrices, suggesting that a distance threshold somewhere between 1.5 and 4 may be appropriate if conducting a season-level analysis to identify the most important parts of attacks across a team's whole season.
The distance threshold was also the most important parameter in terms of determining the run time of MA-Stat-DSM, with a distance threshold of 4 taking nearly five times as long to run per team-match iteration compared to using a distance threshold of 1.5.
Using a high distance threshold (e.g., 20) was found to be computationally prohibitive.

The proposed method could be useful to coaches and performance analysts in basketball who want to identify the most relevant parts of attacks that discriminate between ineffective and effective attacks (or another label if attacks are labeled with something other than effective/ineffective under our proposed definition).
There are potential benefits in utilizing the proposed MA-Stat-DSM method for this purpose rather than watching and coding video in video analysis systems, which is, of course, time-consuming for coaches and performance analysts (and therefore, costly), and may result in biases in the types of play they tend to look for or deem important when conducting post-match reviews.
The MA-Stat-DSM has the automation benefit in that the method can identify aspects of multi-agent behaviour through plays --- and parts of plays in the form of discriminative sub-matrices --- that may not have been obvious to coaches or analysts.
As has been mentioned previously in this paper, an advantage of MA-Stat-DSM is that, unlike machine learning methods, it does not require complex feature engineering of changes in coordinates, player speeds, accelerations, and so on.
Furthermore, the proposed method is more intuitive than deep learning methods, which are generally black-box.
In the current study, the set of possible MA-Stat-DSM parameters needed to be restricted to reduce their number of possible permutations.
It is likely that adjusting the distance threshold upwards and adjusting the significance level upwards (e.g., from 0.05 to 0.1) would have a similar effect in terms of increasing the number of discriminative sub-matrices obtained, however, this could be confirmed in further research.
Other base distances for Hausdorff distance other than Euclidean distance could also be trialed in further work.
Rather than pre-defined areas of the court being used when computing player shooting success probabilities, a more sophisticated approach by partitioning the court could be performed using classification trees as per \cite{zuccolotto2021spatial}.
In general, a limitation of MA-Stat-DSM, despite using a fast implementation of Hausdorff distance, is that its computational complexity was such that it could not be applied to datasets with a very large number of trajectories, e.g., a team's set of attacks from an entire season.
As a result, we needed to consider a team's set of attacks from a single match, and iterate over the team's matches in the season, applying MA-Stat-DSM to team-match datasets.
Future work could seek to improve the speed and scalability of the algorithm.
When visualizing some of the MA-Stat-DSM discriminative sub-matrix results, it would have been useful to have access to the z-coordinate of the ball trajectory, since in some attacks it was not possible to determine whether a pass to avoid a defender was below or above the defender.
%
%
With an appropriate label defined, MA-Stat-DSM could be applied to multi-agent trajectory data from team sports other than basketball, and indeed, to multi-agent trajectory data from domains other than sport.
Finally, the method could be generalized, e.g., such that discriminative sub-matrices do not always need to have a fixed number of agents.

\begin{appendix}
\section*{Appendix}
\subsection*{Code} 
The MA-Stat-DSM code is available on GitHub: 
\url{https://github.com/rorybunker/ma-stat-dsm}

\subsection*{Statistical testing \& pruning properties of (MA)-Stat-DSM}
The statistical testing and pruning properties of MA-Stat-DSM are essentially analogous to the original Stat-DSM \cite{le2020stat}.
Therefore, in this subsection and in Figure 3, we provide a brief explanation of the statistical testing and pruning properties of (MA-)Stat-DSM and refer the reader to \cite{le2020stat} for full details.

Stat-DSM (MA-Stat-DSM) represents sub-trajectories (sub-matrices) in the form of a tree, which is pruned to remove sub-trajectories (sub-matrices) that are guaranteed to not be discriminative (this pruning criterion is shown in line 18 of the MA-Stat-DSM algorithm pseudo-code in Algorithm \ref{alg:algorithm}).
Stat-DSM (MA-Stat-DSM) uses Fisher’s Exact Test (FET) \citep{fisher1922interpretation,agresti1992survey} to determine the statistical significance of a sub-trajectory (sub-matrix) using a contingency table with the number of trajectories (matrices) that, respectively, contain and do not contain sub-trajectories (sub-matrices) within a distance of $\varepsilon$.
A correction for multiple-testing bias is also incorporated, which is necessary due to the calculation of p-values for a large number of trajectories (trajectory matrices), and is conducted using the Westfall-Young (WY) method \citep{westfall1993resampling,terada2013fast}.
Sub-trajectories (sub-matrices) are only identified as SSD if their p-value is less than their adjusted significance level, $\delta$, which is, in turn, less than $\alpha = 0.05$.
The dataset labels are permuted $B = 1000$ times as part of this procedure.
The pruning and WY methods are applied simultaneously to reduce complexity (Step 1, Figure 3).

\end{appendix}

\begin{funding}
This work was supported by JSPS KAKENHI (Grant Numbers 19H04941 and 20H04075) and JST PRESTO (JPMJPR20CA).
\end{funding}

\bibliographystyle{chicago}
\bibliography{main}

\end{document}